\documentclass{article}
\usepackage{amssymb}
\usepackage{amsmath}

\newenvironment{proof}[1][Proof]{\textbf{#1.} }{\ \rule{0.5em}{0.5em}}
\input{tcilatex}

\begin{document}

{\huge \qquad {}A theory about high-temperature}

\qquad \qquad \qquad \qquad {\huge superconductivity}

P.\ Brovetto\footnote{%
Corresponding author E-mail address: brovetto@vaxca1.unica.it.
\par
Tel/Fax: +39-70-6754822}, and V.\ Maxia.

Istituto Nazionale di Fisica della Materia, Sez. Cagliari.

Cittadella universitaria di Monserrato, 09042 Cagliari, Italy.

\bigskip \medskip

\textbf{Abstract}.

We deal with a model for high-temperature superconductivity which maintains
that in cuprates electrons running in the copper oxide layers, found in
lattice of these materials, form spin-singlet bonds with electrons running
in the neighbouring layers. This model reutilizes the BCS\ scheme, but with
the essential difference that the electron pairs are characterized by equal,
rather than opposite momenta as in Cooper pairs. In the present paper, we
consider the electron pair formation and a peculiar canonical transformation
analogous to the transformation once applied to the theory of pairing
correlations in nuclear matter. It is shown that the quasi-particle energy
spectrum remains that of the BCS theory, including the linear relationship
between forbidden energy gap and critical temperature. The model is also
applied to superconductivity of some copperless perovskites of mixed
stoichiometry, whose features are of special worth in understanding the
mechanism of the phenomenon. The possibility of enhancing critical
temperature in cuprates by inserting monovalent ions into the lattice is
considered.



$^{{}}$

{\Large 1. Introduction }

Notwithstanding the great deal of work done till now, no theory about high
temperature superconductivity has obtained a general consensus.\ It is
common belief that even the basic nature of the phenomenon is not
understood.\ In magazine notes appeared on ''Scientific American'' in 2000
and 2004, the inadequacy of theoretical models in explaining superconduction
in cuprates is stressed $\left[ 1\right] .$ Owing to this state of affairs,
we will now examine a mechanism, quite unlike those so far proposed, which
has been conceived by keeping in mind, besides cuprates, the features of
other kinds of unconventional superconductors different from cuprates ($%
\footnote{%
) In Ref. $\left[ 2,\,3,\,4,\,5\right] ,$ some features of this mechanism
have already been presented$.$}$).

Cuprates are surely the most interesting superconductors as they allow for
the highest critical temperatures so far recorded. But a variety of
materials other than cuprates is known, showing a superconductivity not
explained by the BCS theory. Actually, superconduction has been detected in
perovskites of fractional stoichiometry, in mixed copper and alkaly-earth
oxides, in organic compounds and in fullerenes. This makes complex the study
of superconductivity but, at the same time, provides a wide experimental
basis with which theory must be compared. In our opinion, the simplest and
most conservative hypothesis is that the essential superconduction mechanism
is the same in all these materials in spite of their quite different
natures. Accordingly, the very cause of superconductivity must be searched
for in something which is shared by all materials. On that account, next
Section is devoted to singling out the features which pertain to all\
above-mentioned superconductors.\medskip

{\Large 2. Common features of unconventional superconductors }

In reality, two remarkable peculiarities are shared by the superconductors
cited before. The first is that all are characterized by complex layered
lattices or uneven heterogeneous lattices showing discontinuous structures.\
The second is that all contain ions or atoms with unpaired electrons or
electrons not included in closed shells.\ These points are emphasized
hereinafter by considering a selection of various superconductors.

\textit{A) Perovskites with fractional stoichiometries}.\ - Some of these
perovskites are listed in the following Table.

\begin{eqnarray*}
&&\text{Table 1: Superconducting perovskites with fractional stoichiometries.%
} \\
&&\qquad \qquad 
\begin{array}{ccccc}
\text{SrTiO}_{3-\delta }\,\left[ 6\right] & \text{BaPb}_{\,0.7}\text{Bi}%
_{0.3}\text{O}_{3}\,\left[ 7\right] & \text{Ba}_{\,0.6}\text{K}_{0.4}\text{%
BiO}_{3}\,\left[ 8\right] &  &  \\ 
T_{c}\simeq 0.3\,\text{K} & T_{c}=13\,\text{K} & T_{c}=30\,\text{K} &  & 
\end{array}
\\
&&\qquad \qquad 
\begin{array}{ccccc}
&  &  & \text{Sr}_{0.5}\text{K}_{0.5}\text{BiO}_{3}\,\left[ 9\right] & \text{%
Sr}_{0.5}\text{Rb}_{0.5}\text{BiO}_{3}\,\left[ 9\right] \\ 
&  &  & T_{c}=12\,\text{K} & T_{c}=13\,\text{K}%
\end{array}%
\end{eqnarray*}%
Their structure is characterized by lattice discontinuities found at the
borders between cells with different ion compositions.\ The SrTiO$_{3-\delta
}$ superconductor, which shows a partial lack of oxygen, is a reduced
compound.\ But, also BaPb$_{\,0.7}$Bi$_{\,0.3}$O$_{3}$ and Ba$_{\,0.6}$K$%
_{0.4}$BiO$_{3}$ are indeed reduced compounds. In fact, owing to valence
four of lead and five of the fully oxidized bismuth, their stoichiometries
should be written as BaPb$_{0.7}$Bi$_{0.3}$O$_{3.15}$ and Ba$_{0.6}$K$_{0.4}$
BiO$_{3.3},$ respectively. The same argument obviously is right for the
strontium-substituted compounds. For these compounds, the lack of room in
the stiff perovskitic cell prevents oxygen from entering the cell until
metals are fully oxidized. Since oxygen is kept in the form of divalent O$%
^{-2}$ ions, when oxygen is removed as neutral atoms some electrons are left
in the material and become bound to metal ions.\ It follows that unpaired
electrons appear in excess to the noble gas shells of K$^{+1}$, Ba$^{+2},$ Sr%
$^{+2}$ ions or to 5d$^{10}$ shell of Pb$^{+4}$ and Bi$^{+5}$ ions.

\textit{B) Cuprates}.\ - These materials show layered lattices formed by
perovskitic or perovskitic-like cubes. As cuprates are the best known
superconductors, we limit ourselves to few examples.\ The first discovered La%
$_{1.85}$Sr$_{0.15}$CuO$_{4}$ cuprate, which superconducts at 35 K, is
characterized by the K$_{2}$NiF$_{4}$ structure, that is, an alternation of
perovskitic and NaCl-like layers $\left[ 10\right] .$ It shows a fractionary
stoichiometry and is to be regarded as an oxidized superconductor because,
owing to substitution of trivalent lanthanum with divalent strontium, its
stoichiometry should be written as La$_{1.85}$Sr$_{0.15}$CuO$_{3.925}.$ The
92 K superconductor YBa$_{2}$Cu$_{3}$O$_{7}$, usually referred to as YBCO,
is characterized by a stacking of yttrium and barium centred lacunar
perovskitic cubes $\left[ 11\right] .\;$In the so-called TBCO
superconductors, such as for instance the Tl$_{2}$Ba$_{2}$CaCu$_{2}$O$_{8}$
compound, different alternation of perovskitic-like layers of copper,
calcium, barium and thallium oxide are found $\left[ 12,\,13\right] .\;$As
for the unpaired electrons, we point out that all cuprates contain divalent
copper with the $\left[ \text{Ar}\right] $3d$^{9}$ configuration showing
just one unpaired 3d-electron.

\textit{C) Mixed copper and alkaly-earth oxides}. These compounds deserve
attention because they are cuprates lacking in perovskitic structure. The
mixed oxide SrCuO$_{2}$ shows an orthorhombic lattice,\ it is not a
superconductor but superconductivity appears at 40 K in the fractionary
stoichiometry compound Sr$_{0.86}$Nd$_{0.14}$CuO$_{2}$ $\left[ 14\right] .$
It is a reduced compounds because, owing to valence three of neodimium, its
stoichiometry should be written as Sr$_{0.86}$Nd$_{0.14}$CuO$_{2.07}$.\
Apart from the different structure, it is like the superconductors of item 
\textit{A). }Recently, using a field-effect technique, electrons were
removed from (or injected into) the monoclinic CaCuO$_{2}$ compound. In this
way, superconductivity was found at 89 K and 34 K\ depending on wheter 0.15
electrons per molecule are removed or injected, respectively $\left[ 15%
\right] $. Even in this case, superconductivity is originated by
introduction of unpaired electrons and of lattice discontinuities lying at
the borders between cells of different degrees of oxidation (\footnote{%
) With the field-effect technique the average degree of oxidation of the
material can be properly determined. On the contrary, lattice
discontinuities related to the local degree of oxidation of the cells remain
uncertain. Also incidental lattice defects might play a role. \ }).\ 

\textit{D) Organic superconductors}.\ We limit ourselves to the Bechgaard
salt, that is, tetramethyl-tetraselena-fulvalene hexafluoro-phosphate (TMTSF)%
$_{2}$ PF$_{6}$ which superconducts at about 1 K $\left[ 16\right] .\;$This
material is characterized by stackings of strongly bound molecules with a
much weaker intermolecular bonding in direction transverse to the
stackings.\ One electron is moved from one TMTSF\ molecule to one fluorine
atom so that (TMTSF)$^{+1}$ cations and (PF$_{6}$)$^{-1}$ anions appear.
Since in the neutral TMTSF\ molecules all electrons are coupled in $\sigma -$
or $\pi -$bonds, one unpaired electron is present in the (TMTSF)$^{+1}$
cation.

\textit{E) Fullerenes}.\ These materials are characterized by stacks of C$%
_{60}$ balls. Links between carbons in contiguous balls are weaker than
those of carbons in the same ball. Superconductivity has been detected at 18
K in K$_{3}$C$_{60}$ and at 28\ K\ in Rb$_{3}$C$_{60}$ $\left[ 17\right] .$
The presence of potassium or rubidium atoms, showing the $\left[ \text{Ar}%
\right] $4s and $\left[ \text{Kr}\right] $5s configurations, respectively,
inserts unpaired electrons in the C$_{60}$ stacks. With the field-effect
technique, electrons were removed or injected into the C$_{60}$ balls, so
leaving some unpaired electrons there.\ In this way, superconductivity was
originated at peak temperatures of 52\ K or 11 K when just three electrons
were taken off or added to each ball, respectively $\left[ 18\right] .$

The previous analysis confirms that uneven lattices and unpaired electrons
are really features common to the superconductors considered above. However,
different kinds of superconductors must be distinguished depending on the
actual provenance of the unpaired electrons. Indeed, compounds of item 
\textit{A) }and the Sr$_{0.86}$Nd$_{0.14}$CuO$_{2}$ compound are
\textquotedblright reduced\textquotedblright\ superconductors.\ The La$%
_{1.85}$Sr$_{0.15}$CuO$_{4}$ cuprate, on the contrary, is an
\textquotedblright oxidized\textquotedblright\ superconductor. Cuprates as
YBa$_{2}$Cu$_{3}$O$_{7}$ or Tl$_{2}$Ba$_{2}$CaCu$_{2}$O$_{8}$ and the
Bechgaard salt are to be regarded as \textquotedblright
intrinsic\textquotedblright\ superconductors, since unpaired electrons are
peculiar to their chemical composition. The alkaly-doped fullerene as well
as fullerene and CaCuO$_{2}$ oxide showing field-effect superconductivity
are to be regarded as \textquotedblright doped\textquotedblright\
superconductors, because in these materials superconductivity is originated
by an external agent.

In our opinion,\ the features examined above can be considered as a sort of\
\textquotedblright Rosetta stone\textquotedblright\ for disentangling the
problem of high temperature superconductivity.\medskip

{\Large 3. About properties of fermion systems }

Let us recall some topics concerning properties of fermion systems which
will be helpful in understanding the mechanism of superconduction in the
materials cited above. In 1916, G.N. Lewis first discovered that covalent
bonds consist of pairs of shared electrons $\left[ 19\right] $. This fact,
inexplicable by the classical physics, was interpreted in 1927 by W. Heitler
and F.\ London (HL) who applied quantum mechanics to the hydrogen molecule $%
\left[ 20\right] $. By considering two hydrogen atoms $A$ and $B$ in 1s
states, they wrote a two-electron wave function of the form: $\left[
u_{1sA}\left( 1\right) u_{1sB}\left( 2\right) +u_{1sB}\left( 1\right)
u_{1sA}\left( 2\right) \right] $ in which each electron is found at the same
time both on atom $A$ and $B$. This function, symmetric with respect to
exchange of electrons, was associated to an antisymmetric spin function: $%
\left[ \alpha \left( 1\right) \beta \left( 2\right) -\beta \left( 1\right)
\alpha \left( 2\right) \right] $ representing a spin-singlet state, so
allowing for the Pauli principle. In this way, in evaluating the expectation
value of energy, integrals involving products of electron states: $%
u_{1sA}\left( 1\right) u_{1sB}\left( 1\right) $ and $u_{1sA}\left( 2\right)
u_{1sB}\left( 2\right) $ appear in calculations. These exchange integrals
account for covalent bond energy.\ A year later, W.\ Heisemberg, utilizing
the same arguments, explained the origin of the Weiss field in ferromagnetic
solids $\left[ 21\right] $.

In 1933, exchange forces came back into evidence in a quite different field
of physics. In this year, indeed, E.\ Majorana, in dealing with nuclear
interactions, introduced forces which exchange the coordinates of the
interacting nucleons $\left[ 22\right] $.\ These forces are mediated by
charged pions and act only for nucleons in neighbouring momentum states $%
\left[ 23\right] $.\ The Majorana forces are of paramount importance since
they account for the saturation effects in binding energy of nuclei.

In 1957\ the famous BCS theory finally explained superconduction in metals $%
\left[ 24\right] $. The essential device of this theory are the Cooper
pairs, that is, pairs of electrons of opposite momenta bound by a phonon
coupling. Utilizing a special canonical transformation devised by N.N.\
Bogolyubov, the system of interacting electrons is substituted by a set of
non-interacting quasi-particles showing an energy gap at the top of the
distribution $\left[ 25\right] $.\ 

The great success of the BCS\ theory drew attention on the possibility of
its application to nuclear physics. In 1958, A.\ Bohr, B.R.\ Mottelson and
D.\ Pines proposed that the energy gap found in the spectra of even-even
nuclei is originated by a mechanism analogous to that of superconduction in
metals $\left[ 26\right] $.\ Indeed, when the main part of the
nucleon-nucleon interactions is averaged so allowing for a self-consistent
field, there remains a residual weak interaction, due to Majorana forces,
which couples nucleons of like momenta. A thorough treatment of this problem
was performed by S.T.\ Belyaev which modified the Bogolyubov transformation
substituting the Cooper pairs with Majorana pairs of nucleons with equal
linear momenta, but opposite projections of angular momenta along the
quantization axis $\left[ 27\right] .\;$This treatment, however, leaves out
the dependence on temperature of the energy gap, owing to the fact that
nuclei are always on the ground state. An equivalent treatment was performed
by L.P.\ Gor'kov and A.I.\ Alekseev utilizing the Green function technique $%
\left[ 28,\,29\right] $.

Since our treatment on superconductivity utilizes a method similar to that
applied by Belyaev, an extensive account on this matter is given in Appendix
A.\medskip

{\Large 4. The superconducting Lewis pairs}

The superconductor features highlighted in Section 2 and the arguments
presented in Section 3 induce us to argue that at low temperature unpaired
electrons running in a superconductor region bordering on a lattice
discontinuity originate spin-singlet pairs with electrons running in the
region bordering on the opposite side of the lattice discontinuity. This is
due to instability of the unpaired electrons that tend to form covalent
bonds. Obviously, in order to set out a quantitative treatment, it is
necessary to know, besides the electron wave functions, the actual nature of
the lattice discontinuities. This occurs with the intrinsic superconductors,
such as the the YBa$_{2}$Cu$_{3}$O$_{7}\ $and the Tl$_{2}$Ba$_{2}$CaCu$_{2}$O%
$_{8}$ cuprates, or the Bechgaard salt. On the contrary, with the reduced
superconductors of items \textit{A)} and \textit{C)} it is necessary to
resort to special conjectures, since their fractional stoichiometries make
the structure uncertain. These superconductors are investigated in Section
7. Also the fullerene-based supercondutors give rise to difficulties of this
kind. In practice, the materials most right for our investigations are the
intrinsic cuprates.

The previously cited cuprates, are characterized by planes of oxygen lacunae
and yttrium or calcium ions sandwiched between couples of contiguous CuO$%
_{2} $ layers (see $\left[ 30\right] $ Ch. 7). In the following, these
layers will be marked with labels $a$ and $b$. Two unpaired electrons, one
running on layer $a$ the other on layer $b$, can be represented by the
tight-binding (TB) wave functions

\begin{equation*}
\phi _{a}(\mathbf{k}_{a},\mathbf{r}_{1})=\frac{1}{\sqrt{N}}%
\sum_{p=1}^{N}\exp \left( i\,\mathbf{k}_{a}\cdot \mathbf{u}_{p}\right) \,\,a(%
\mathbf{r}_{1}-\mathbf{u}_{p}),
\end{equation*}

\begin{equation}
\phi _{b}(\mathbf{k}_{b},\mathbf{r}_{2})=\frac{1}{\sqrt{N}}%
\sum_{q=1}^{N}\exp \left( i\,\mathbf{k}_{b}\cdot \mathbf{v}_{q}\right) \,\,b(%
\mathbf{r}_{2}-\mathbf{v}_{q}),  \label{pat}
\end{equation}%
in which $\mathbf{k}_{a}\ $and $\mathbf{k}_{b}$ mean the electron wave
vectors, $a(\mathbf{r}_{1}-\mathbf{u}_{p})$ and $b(\mathbf{r}_{2}-\mathbf{v}%
_{q})$ the 3d-orbitals of the copper ions on layers $a$ and $b$ and $\mathbf{%
u}_{p}$ and $\mathbf{v}_{q}$ their lattice vectors, respectively. Each
copper ion on layer $a$ is separated from a corresponding ion on layer $b$
by the spacing $\mathbf{\lambda }$ between the layers, that is,%
\begin{equation}
\mathbf{v}_{p}-\mathbf{u}_{p}=\mathbf{\lambda .}  \label{Aa}
\end{equation}%
The energies of the unpaired electrons spoken of are

\begin{equation*}
W_{a}\left( \mathbf{k}_{a}\right) =\left\langle \phi _{a}(\mathbf{k}_{a},%
\mathbf{r}_{1})\right\vert H_{a}\left( \mathbf{p}_{1},\mathbf{r}_{1}\right)
\left\vert \phi _{a}(\mathbf{k}_{a},\mathbf{r}_{1})\right\rangle ,
\end{equation*}

\begin{equation}
W_{b}\left( \mathbf{k}_{b}\right) =\left\langle \phi _{b}(\mathbf{k}_{b},%
\mathbf{r}_{2})\right\vert H_{b}\left( \mathbf{p}_{2},\mathbf{r}_{2}\right)
\left\vert \phi _{b}(\mathbf{k}_{b},\mathbf{r}_{2})\right\rangle ,
\label{zzz}
\end{equation}%
where the Hamiltonians $H_{a}$ and $H_{b}$ account for the electron kinetic
energies $\mathbf{p}_{1}^{2}/2m$ and $\mathbf{p}_{2}^{2}/2m$ and for the
Coulomb interactions of electrons $1$ and $2$ with the copper ions in
positions $\mathbf{u}_{p}$ and $\mathbf{v}_{q},$ respectively.\ For $\mathbf{%
k}_{a}=\mathbf{k}_{b}$ , energies $W_{a}\left( \mathbf{k}_{a}\right) $ and $%
W_{b}\left( \mathbf{k}_{b}\right) $ are equal, owing to equality of the CuO$%
_{2}$ layers. In Appendix B, utilizing a special model for the actual nature
of the copper ion orbitals, energies $W_{a}\left( \mathbf{k}_{a}\right) $
and $W_{b}\left( \mathbf{k}_{b}\right) $ are evaluated on the ground of Eqs.
(\ref{zzz}).

Owing to the peculiar structure of the before cited cuprates, that is, the
presence of oxygen lacunae on the yttrium or calcium planes placed between
the copper ions, an unpaired electron of layer $a$ is allowed to form a
covalent bond with an unpaired electron of layer $b,$ like the 1s electron
of a hydrogen atom $A$ forms a covalent bond with the 1s electron of another
hydrogen atom $B$. On this ground, in analogy to the HL treatment of the
hydrogen molecule $\left[ 31\right] $, the wave function for a pair of
electrons of layers $a$ and $b$ in a spin-singlet state is

\begin{equation*}
\Psi \left( \mathbf{r}_{1},\,\mathbf{r}_{2}\right) =\frac{1}{\sqrt{2\left(
1+\langle \phi _{a}\mid \phi _{b}\rangle ^{2}\right) }}\left[ \phi _{a}(%
\mathbf{k}_{a},\mathbf{r}_{1})\phi _{b}(\mathbf{k}_{b},\mathbf{r}_{2})+\phi
_{a}(\mathbf{k}_{a},\mathbf{r}_{2})\phi _{b}(\mathbf{k}_{b},\mathbf{r}_{1})%
\right] \times
\end{equation*}

\begin{equation}
\times \frac{1}{\sqrt{2}}\left[ \alpha \left( 1\right) \beta \left( 2\right)
-\alpha \left( 2\right) \beta \left( 1\right) \right] ,  \label{pet}
\end{equation}%
$\alpha \left( 1\right) $ and $\beta \left( 2\right) $ standing for the spin
functions. In the following, these pairs are referred to as
\textquotedblright Lewis pairs\textquotedblright\ since this author, already
cited in Section 3, pioneered investigations on covalent bonds. The
possibility of applying the HL treatment is due to the fact that it is
implemented aside from the actual nature of the electron states, so that 1s$%
_{A}$ and 1s$_{B}$ or $\phi _{a}$ and $\phi _{b}$ states can be
indifferently considered. This notwithstanding the fact that 1s states
account for a single Coulomb potential centre, while $\phi $ states account
for $N$ centres. Like in the HL treatment, energy $-W_{P}$ of the electron
pair holds a \textquotedblright classic\textquotedblright\ contribution,
that is, without exchange of electrons between $\phi _{a}$ and $\phi _{b}$
states and an exchange contribution in which both electrons are shared
between $\phi _{a}$ and $\phi _{b}$ states (\footnote{%
) Apart from substitution of 1s$_{A}$ and 1s$_{B}$ hydrogen-like states with 
$\phi _{a}$ and $\phi _{b}$ states and the presence of summations over the $%
N $ copper ions, the terms appearing in Eq. (5) are like the terms in Eq.
(43-7) and (43-9) of reference $\left[ 31\right] $ p. 342. We omit
considerig terms for the spin-triplet state which, in HL treatment,
originate repulsion between the hydrogen atoms.}), that is,%
\begin{equation*}
-W_{P}=\left\langle \Psi \left( \mathbf{r}_{1},\mathbf{r}_{2}\right)
\right\vert H_{int}\left( \mathbf{r}_{1},\mathbf{r}_{2}\right) \left\vert
\Psi \left( \mathbf{r}_{1},\mathbf{r}_{2}\right) \right\rangle =
\end{equation*}

\begin{equation*}
=\frac{\left\langle \phi _{a}\left( \mathbf{k}_{a},\mathbf{r}_{1}\right)
\phi _{b}\left( \mathbf{k}_{b},\mathbf{r}_{2}\right) \right\vert }{\sqrt{%
1+\left\langle \phi _{a}\mid \phi _{b}\right\rangle ^{2}}}\left[
-\sum_{q=1}^{N}\frac{Ze^{2}}{\left\vert \mathbf{r}_{1}-\mathbf{v}%
_{q}\right\vert }-\right.
\end{equation*}

\begin{equation*}
-\left. \sum_{p=1}^{N}\frac{Ze^{2}}{\left\vert \mathbf{r}_{2}-\mathbf{u}%
_{p}\right\vert }+\frac{e^{2}}{r_{1,2}}\right] \frac{\left\vert \phi
_{a}\left( \mathbf{k}_{a},\mathbf{r}_{1}\right) \phi _{b}\left( \mathbf{k}%
_{b},\mathbf{r}_{2}\right) \right\rangle }{\sqrt{1+\left\langle \phi
_{a}\mid \phi _{b}\right\rangle ^{2}}}+
\end{equation*}

\begin{equation*}
+\frac{\left\langle \phi _{a}\left( \mathbf{k}_{a},\mathbf{r}_{1}\right)
\phi _{b}\left( \mathbf{k}_{b},\mathbf{r}_{2}\right) \right\vert }{\sqrt{%
1+\left\langle \phi _{a}\mid \phi _{b}\right\rangle ^{2}}}\left[
-\sum_{p=1}^{N}\frac{Ze^{2}}{\left\vert \mathbf{r}_{1}-\mathbf{u}%
_{p}\right\vert }-\right.
\end{equation*}

\begin{equation}
-\left. \sum_{q=1}^{N}\frac{Ze^{2}}{\left\vert \mathbf{r}_{2}-\mathbf{v}%
_{q}\right\vert }+\frac{e^{2}}{r_{1,2}}\right] \frac{\left\vert \phi
_{b}\left( \mathbf{k}_{b},\mathbf{r}_{1}\right) \phi _{a}\left( \mathbf{k}%
_{a},\mathbf{r}_{2}\right) \right\rangle }{\sqrt{1+\left\langle \phi
_{a}\mid \phi _{b}\right\rangle ^{2}}}.  \label{frf}
\end{equation}

In this equation, Hamiltonian $H_{int}\left( \mathbf{r}_{1},\mathbf{r}%
_{2}\right) $ allows$\ $for Coulomb interactions of electrons with copper
ions of effective charge $Z$ and for Coulomb repulsion between the
electrons. Like in the HL treatment, it follows that pairing energy is given
by%
\begin{equation}
-W_{P}=\frac{2J+J^{\prime }+2\langle \phi _{a}\mid \phi _{b}\rangle
K+K^{\prime }}{1+\langle \phi _{a}\mid \phi _{b}\rangle ^{2}}\simeq 2\langle
\phi _{a}\mid \phi _{b}\rangle K+K^{\prime }.  \label{pot}
\end{equation}%
Terms $J$ and $J^{\prime }$ represent the classic contributions to energy of
the electron pair. These contributions are negligible as occurs in the case
of the hydrogen molecule $\left[ 31\right] $. The squared overlap integral $%
\langle \phi _{a}\mid \phi _{b}\rangle ^{2}$ is negligible with respect to
unity. Only integrals $K$ and $K^{\prime },$ which account for exchange
interactions, must be retained.

We proceed now to evaluate the integrals appearing in Eq. (\ref{pot}). Let
us first consider the exchange integral $K^{\prime }.$ Taking into account
Eqs. (\ref{pat}), we have

\begin{equation*}
K^{\prime }=\left\langle \phi _{a}\left( \mathbf{k}_{a},\mathbf{r}%
_{1}\right) \phi _{b}\left( \mathbf{k}_{b},\mathbf{r}_{2}\right) \right| 
\frac{e^{2}}{r_{1,2}} \left| \phi _{b}\left( \mathbf{k}_{b},\mathbf{r}%
_{1}\right) \phi _{a}\left( \mathbf{k} _{a},\mathbf{r}_{2}\right)
\right\rangle =
\end{equation*}

\begin{equation*}
=\frac{e^{2}}{N^{2}}\int \sum_{q,\,p=1}^{N}\exp \left( -i\mathbf{k}_{b}\cdot 
\mathbf{v}_{q}+i\mathbf{k}_{a}\cdot \mathbf{u}_{p}\right) b\left( \mathbf{r}%
_{2}-\mathbf{v} _{q}\right) a\left( \mathbf{r}_{2}-\mathbf{u}_{p}\right)
\times
\end{equation*}

\begin{equation}
\times \left[ \int \sum_{p,\,q=1}^{N}\exp \left( -i\mathbf{k}_{a}\cdot 
\mathbf{u}_{p}+i\mathbf{k}_{b}\cdot \mathbf{v}_{q}\right) \frac{a\left( 
\mathbf{r}_{1}-\mathbf{u}_{p}\right) b\left( \mathbf{r}_{1}-\mathbf{v}%
_{q}\right) }{r_{1,\,2}}d^{3}\mathbf{r}_{1}\right] d^{3}\mathbf{r}_{2}.
\label{ooo}
\end{equation}%
This equation involves sums over $N^{4}$ terms. But, taking into account
that the electron distributions in orbitals $a(\mathbf{r}_{1}-\mathbf{u}%
_{p}) $ or $a(\mathbf{r}_{2}-\mathbf{u}_{p})$ and $b(\mathbf{r}_{1}-\mathbf{v%
}_{q}) $ or $b(\mathbf{r}_{2}-\mathbf{v}_{q})$ are closely localized at the
lattice positions $\mathbf{u}_{p}$ and $\mathbf{v}_{q},$ respectively, terms 
$b\left( \mathbf{r}_{2}-\mathbf{v}_{q}\right) a\left( \mathbf{r}_{2}-\mathbf{%
u}_{p}\right) $ and $a\left( \mathbf{r}_{1}-\mathbf{u}_{p}\right) b\left( 
\mathbf{r}_{1}-\mathbf{v}_{q}\right) $ are non-negligible only when $\mathbf{%
r}_{2}\simeq \mathbf{v}_{q}$ and $\mathbf{r}_{2}\simeq \mathbf{u}_{p}$ and $%
\mathbf{r}_{1}\simeq \mathbf{u}_{p}$ and $\mathbf{r}_{1}\simeq \mathbf{v}%
_{q},$ which entails that $\mathbf{v}_{q}\simeq \mathbf{u}_{p},$ that is,
remembering Eq. (\ref{Aa}), $p=q.$ It follows that the sums in Eq. (\ref{ooo}%
) contain only $N^{2}$ non-negligible terms. This leads to

\begin{equation*}
K^{\prime }=\frac{e^{2}}{N^{2}}\int \sum_{q,\,p=1}^{N}\exp \left[ i\left( 
\mathbf{k}_{a}-\mathbf{k}_{b}\right) \left( \mathbf{v}_{q}-\mathbf{u}_{p}-%
\mathbf{\lambda }\right) \right] \times
\end{equation*}

\begin{equation}
\times \frac{b\left( \mathbf{r}_{2}-\mathbf{v}_{q}\right) a\left( \mathbf{r}%
_{2}-\mathbf{u}_{q}\right) a\left( \mathbf{r}_{1}-\mathbf{u}_{p}\right)
b\left( \mathbf{r}_{1}-\mathbf{v}_{p}\right) }{r_{1,\,2}}d^{3}\mathbf{r}%
_{1}d^{3}\mathbf{r}_{2}.  \label{rrr}
\end{equation}%
In this sum, significant contributions appear only when $r_{1,\,2}\ $is
small, that is when $\mathbf{r}_{2}\simeq \mathbf{r}_{1}$ (\footnote{%
) It is to be pointed out that Coulomb repulsion between electrons is
screened by the presence of the positive ions which assure the electric
neutrality of the material. Consequently, Coulomb repulsion decreases more
quickly than $e^{2}/r_{1,2}$. This is like what occurs with Cooper pairs in
BCS\ theory.}). This, as before, entails that only terms with $\mathbf{v}%
_{q}\simeq \mathbf{u}_{p}$,$\ $that is, $p=q$ are non-negligible.\ So we
obtain

\begin{equation*}
K^{\prime }=\frac{e^{2}}{N^{2}}\int \sum_{q=1}^{N}\frac{b\left( \mathbf{r}%
_{2}-\mathbf{v}_{q}\right) a\left( \mathbf{r}_{2}-\mathbf{u}_{q}\right)
a\left( \mathbf{r}_{1}-\mathbf{u}_{q}\right) b\left( \mathbf{r}_{1}-\mathbf{v%
}_{q}\right) }{r_{1,\,2}}d^{3}\mathbf{r}_{1}d^{3}\mathbf{r}_{2}=
\end{equation*}

\begin{equation}
=O\left( \frac{1}{N}\right) ,  \label{sss}
\end{equation}%
which means that this integral, which accounts for Coulomb repulsion between
electrons, can be neglected. We consider now the overlap integral

\begin{equation*}
\langle \phi _{a}\left( \mathbf{k}_{a},\mathbf{r}_{1}\right) \mid \phi
_{b}\left( \mathbf{k}_{b},\mathbf{r}_{1}\right) \rangle =
\end{equation*}
\begin{equation}
=\frac{1}{N}\int \sum_{p,\,q=1}^{N}\exp \left( -i\mathbf{k}_{a}\cdot \mathbf{%
u}_{p}+i\mathbf{k}_{b}\cdot \mathbf{v}_{q}\right) a\left( \mathbf{r}_{1}-%
\mathbf{u}_{p}\right) b\left( \mathbf{r}_{1}-\mathbf{v}_{q}\right) d^{3}%
\mathbf{r}_{1}.  \label{ioi}
\end{equation}%
In this case, significant contributions are obtained only for orbitals with $%
\mathbf{u}_{p}\simeq \mathbf{r}_{1}$ and $\mathbf{v}_{q}\simeq \mathbf{r}%
_{1} $ which entails that $\mathbf{u}_{p}\simeq $ $\mathbf{v}_{q}$ and, as
before, $p=q.$ We have thus

\begin{equation*}
\langle \phi _{a}\left( \mathbf{k}_{a},\mathbf{r}_{1}\right) \mid \phi
_{b}\left( \mathbf{k}_{b},\mathbf{r}_{1}\right) \rangle =
\end{equation*}

\begin{equation}
=\int \exp \left[ -i\left( \mathbf{k}_{a}-\mathbf{k}_{b}\right) \cdot 
\mathbf{r}_{\shortparallel \,1}^{{}}\right] \frac{1}{N}\sum_{p=1}^{N}a\left( 
\mathbf{r}_{1}-\mathbf{u}_{p}\right) b\left( \mathbf{r}_{1}-\mathbf{u}_{p}-%
\mathbf{\lambda }\right) d^{3}\mathbf{r}_{1},  \label{AA}
\end{equation}%
in which only the parallel component $\mathbf{r}_{\shortparallel \,1}^{{}}$
has been accounted for in the exponential factor since $\mathbf{k}_{a}$ and $%
\mathbf{k}_{b}$ are parallel to $a$ and $b$ layers. By putting

\begin{equation}
F\left( \mathbf{r}_{\mathbf{\shortparallel }1}\right) =\frac{1}{N}
\sum_{p=1}^{N}\int a\left( \mathbf{r}_{1}-\mathbf{u}_{p}\right) b\left( 
\mathbf{r} _{1}-\mathbf{u}_{p}-\mathbf{\lambda }\right) \,dr_{\perp \,1},
\label{xyz}
\end{equation}
Eq. (\ref{AA}) can be rewritten as a Fourier transform, that is, 
\begin{equation}
\langle \phi _{a}\left( \mathbf{k}_{a},\mathbf{r}_{1}\right) \mid \phi
_{b}\left( \mathbf{k}_{b},\mathbf{r}_{1}\right) \rangle =\int \exp \left[
-i\,\left( \mathbf{k} _{a}-\mathbf{k}_{b}\right) \cdot \mathbf{r}_{\mathbf{%
\shortparallel }1}\right] \,F\left( \mathbf{r}_{\mathbf{\shortparallel }%
1}\right) \,d^{2}\mathbf{r}_{\mathbf{\shortparallel }1}.  \label{xzy}
\end{equation}
Function $F\left( \mathbf{r}_{\mathbf{\shortparallel }1}\right) $ shows a
slight periodic dependence on $\mathbf{r}_{\mathbf{\shortparallel }1}$
since, when $\mathbf{r}_{\mathbf{\shortparallel }1}$ varies, about
equivalent terms are always included in the sum over $p$. In particular, if
dependence on $\mathbf{r}_{\mathbf{\shortparallel }1}$ of function $F\left( 
\mathbf{r}_{\mathbf{\shortparallel } 1}\right) $ is omitted we have

\begin{equation}
\langle \phi _{a}\left( \mathbf{k}_{a},\mathbf{r}_{1}\right) \mid \phi
_{b}\left( \mathbf{k}_{b},\mathbf{r}_{1}\right) \rangle =\left( 2\pi \right)
^{2}F\,\delta \left( \mathbf{k}_{a}-\mathbf{k}_{b}\right) .  \label{zyx}
\end{equation}
Eqs. (\ref{xzy}) and (\ref{zyx}) entail that electrons of equal wave
vectors, that is, of equal momenta, allow for the maximum value of overlap
integral and, therefore, the maximum pairing energy. It is worth to point
out that this result mimics that concerning the Majorana exchange forces,
which likewise originate fermion pairs of equal momenta. But, with the Lewis
pairs this is a consequence of antisymmetry of the wave function (\ref{pet})
which accounts for the Pauli's principle, while with the Majorana pairs
equality of momenta directly follows from the exchange nature of the forces,
which are mediated by charged pions. On this ground, we consider in the
following only electron pairs with $\mathbf{k}_{a}=\mathbf{k}_{b}.$ We find
in this way

\begin{equation}
\langle \phi _{a}\left( \mathbf{k}_{a},\mathbf{r}_{1}\right) \mid \phi
_{b}\left( \mathbf{k}_{b},\mathbf{r}_{1}\right) \rangle =\frac{1}{N}%
\,\sum_{p=1}^{N}\int a(\mathbf{r}_{1}-\mathbf{u}_{p})\,b(\mathbf{r}_{1}-%
\mathbf{u}_{p}-\mathbf{\lambda })d^{\,3}\mathbf{r}_{1}=S_{a,b},  \label{sac}
\end{equation}
where

\begin{equation}
S_{a,b}=\int a(\mathbf{\rho })\,b(\mathbf{\rho }-\mathbf{\lambda })\,d^{\,3}%
\mathbf{\rho }.  \label{sec}
\end{equation}
By applying the same procedure to exchange integral

\begin{equation}
K=-\left\langle \phi _{b}(\mathbf{k}_{b},\mathbf{r}_{2})\right|
\sum_{p=1}^{N}\frac{Z\,e^{2}}{\left| \mathbf{r}_{2}-\mathbf{u}_{p}\right| }%
\left| \phi _{a}(\mathbf{k}_{a},\mathbf{r}_{2})\right\rangle ,  \label{sic}
\end{equation}
we obtain

\begin{equation}
K=-\frac{1}{N}\sum_{l=1}^{N}\int a(\mathbf{r}_{2}-\mathbf{u}_{l})\frac{%
Z\,e^{2}}{ \mid \mathbf{r}_{2}-\mathbf{u}_{l}\mid }b(\mathbf{r}_{2}-\mathbf{u%
}_{l}-\mathbf{\lambda } )d^{\,3}\mathbf{r}_{2}=E_{b,\,a},  \label{soc}
\end{equation}
where

\begin{equation}
E_{b,\,a}=-\int a(\mathbf{\rho })\frac{Z\,e^{2}}{\mid \mathbf{\rho }\mid }b(%
\mathbf{\rho }-\mathbf{\lambda })d^{\,3}\mathbf{\rho }.  \label{suc}
\end{equation}
Consequently, utilizing Eqs. (\ref{pot}), (\ref{sss}), (\ref{sac}) and (\ref%
{soc}), pairing energy for $\mathbf{k}_{a}=\mathbf{k}_{b}$ turns out to be

\begin{equation}
-W_{P}=2\,S_{a,\,b}\,E_{b,\,a}.  \label{ser}
\end{equation}%
Energy $W_{P}$, of course, is expected to be very small in comparison with
those of most covalent bonds, owing to the considerable separation of the
interacting copper oxide layers which is larger than normal covalent bond
lengths.

\medskip

{\Large 5. The canonical transformation}

According to the above picture, electrons running on the contiguous copper
oxide layers $a$ and $b$ constitute one set of $2N$ Fermi's particles. An
electron in this set with a given wave-vector $\mathbf{k}$ shows
degeneration due to its spin components $\alpha $ and $\beta $ and
degeneration due to options $\mathbf{k=k}_{a}$ or $\mathbf{k}=\mathbf{k}%
_{b}, $ that is, to its placing on layer $a$ or $b$. Owing to $\mathbf{k}$
degeneration, the electron set is split into two conjugated subsets $a$ and $%
b$. Each electron of subset $a$ is paired with an electron of subset $b$ of
equal wave vector, that is, of equal momentum, with an energy $-W_{P}$
independent of the actual momentum. As follows from spin-singlet wave
function (\ref{pet}), these pairs show spin zero so that can be regarded as
weakly bound bosons$.$ On this ground, the second quantization Hamiltonian
for Lewis pairs can be identified with the Belyaev's Hamiltonian (\ref{xhz})
considered in Appendix A in connection with systems of interacting particles
characterized by symmetry of reflection in a plane. This, of course, barring
substitution of indexes $R$ and $L$ with indexes $a$ and $b$. In the case of
YBa$_{2}$Cu$_{3}$O$_{7}$ cuprate, the plane of symmetry is to be identified
with the plane of yttrium ions. In this way, by taking single particle
energies into account, writing briefly $\varepsilon _{k}$ for $W_{a}\left( 
\mathbf{k}_{a}\right) $ and $W_{b}\left( \mathbf{k}_{b}\right) $ and
substituting $V_{k,k^{\prime }}$ with $W_{P}$, the Hamiltonian for a system
of electrons forming Lewis pairs can be written in the quick form

\begin{equation}
\widehat{H}=\sum_{k}\left( \varepsilon _{k}-\mu \right) \left( \widehat{%
\alpha }_{ka}^{+}\widehat{\alpha }_{ka}^{{}}+\widehat{\alpha }_{kb}^{+}%
\widehat{\alpha }_{kb}^{{}}\right) -W_{P}\sum_{k,k^{\prime }}\widehat{\alpha 
}_{k^{\prime }a}^{+}\widehat{\alpha }_{k^{\prime }b}^{+}\widehat{\alpha }%
_{kb}^{{}}\widehat{\alpha }_{ka}^{{}}  \label{bim}
\end{equation}%
$\mu $ standing for the chemical potential of the $2N$ electron set. In
Appendix C, simple arguments are posed showing that this Hamiltonian
conserves the momentum of the system.

Assuming coefficients which satisfy condition

\begin{equation}
U_{k}^{2}+V_{k}^{2}=1,  \label{bom}
\end{equation}
the canonical transformation is

\begin{equation}
\begin{array}{c}
\widehat{\alpha }_{ka}=U_{k}\widehat{\,\beta }_{ka}+V_{k}\widehat{\,\beta }%
_{kb}^{+}, \\ 
\widehat{\alpha }_{kb}=U_{k}\widehat{\,\beta }_{kb}-V_{k}\widehat{\,\beta }%
_{ka}^{+},%
\end{array}
\label{bum}
\end{equation}%
$\widehat{\,\beta }_{ka}$ and $\widehat{\,\beta }_{kb}$ standing for the
quasi-particle destruction operators. Substitution of Eq. (\ref{bum}) into
Eq. (\ref{bim}) leads to

\begin{equation*}
\widehat{H}=\sum_{k}\xi _{k}\left[ 2V_{k}^{2}+\left(
U_{k}^{2}-V_{k}^{2}\right) \left( \widehat{n}_{ka}+\widehat{n}_{kb}\right)
+2U_{k}V_{k}\left( \widehat{\beta }_{ka}^{+}\widehat{\beta }_{kb}^{+}+%
\widehat{\beta }_{kb}^{{}}\widehat{\beta }_{ka}^{{}}\right) \right] -
\end{equation*}

\begin{equation}
-W_{P}\sum_{k,\,k^{\prime }}\widehat{B}_{k^{\prime }}^{+}\widehat{B}%
_{k}^{{}},  \label{cam}
\end{equation}%
where $\xi _{k}=\varepsilon _{k}-\mu $ and 
\begin{equation}
\widehat{B}_{k}=U_{k}^{2}\widehat{\beta }_{kb}^{{}}\widehat{\beta }%
_{ka}^{{}}-V_{k}^{2}\widehat{\beta }_{ka}^{+}\widehat{\beta }%
_{kb}^{+}+U_{k}V_{k}\left( 1-\widehat{n}_{ka}-\widehat{n}_{kb}\right) ,
\label{cem}
\end{equation}%
$\widehat{n}_{ka}=\widehat{\beta }_{ka}^{+}\widehat{\beta }_{ka}^{{}}$ and $%
\widehat{n}_{kb}=\widehat{\beta }_{kb}^{+}\widehat{\beta }_{kb}^{{}}$
standing for the operators of quasi-particle numbers. These equations, apart
from some obvious differences in symbols$,$ are like those for the BCS
theory given, for instance, in the Landau and Lifshitz treatise $\left[ 32%
\right] $. For this reason, reutilizing the same routine procedure, we limit
ourselves to reporting the most significant issues. By taking condition (\ref%
{bom}) into account, considering the diagonal terms in the Hamiltonian and
minimizing energy with respect to coefficient $U_{k}$ for given entropy, we
find

\begin{equation}
2\xi _{k}U_{k}V_{k}=\Delta \left( U_{k}^{2}-V_{k}^{2}\right) ,  \label{cim}
\end{equation}
with

\begin{equation}
\Delta =W_{P}\sum_{k^{\prime }}U_{k^{\prime }}V_{k^{\prime }}\left(
1-n_{k^{\prime }a}-n_{k^{\prime }b}\right) ,  \label{com}
\end{equation}%
in which $n_{k^{\prime }a}\ $and $n_{k^{\prime }b}$ now mean the actual
numbers of quasi-particles. When Eq. (\ref{cim}) is verified, the
non-diagonal second-order terms $\widehat{\beta }_{kb}\widehat{\beta }_{ka}$
and $\widehat{\beta }_{ka}^{+}\widehat{\beta }_{kb}^{+}$ are removed from
Hamiltonian (\ref{cam}). From Eqs. (\ref{bom}) and (\ref{cim}), the usual
relations for coefficients $U_{k}$ and $V_{k}$ are obtained,

\begin{equation}
U_{k}^{2}=\frac{1}{2}\left( 1+\frac{\xi _{k}}{\sqrt{\Delta ^{2}+\xi _{k}^{2}}%
}\right) ,\,\,\,\,\,\,\,\,\,\,\,\,\,\,\,\,\,\,\,\,\,\,\,\,\,\,V_{k}^{2}=%
\frac{1}{2}\left( 1-\frac{\xi _{k}}{\sqrt{\Delta ^{2}+\xi _{k}^{2}}}\right) ,
\label{cum}
\end{equation}%
which, when substituted into Eq. (\ref{com}), lead to the equation which
determines $\Delta $

\begin{equation}
W_{P}\sum_{k^{\prime }}\frac{\left( 1-n_{k^{\prime }a}-n_{k^{\prime
}b}\right) }{\sqrt{\Delta ^{2}+\xi _{k^{\prime }}^{2}}}=1.  \label{dam}
\end{equation}

The next step is to change summation on $k^{\prime }$ to integration on
energy. Taking into account that for $\varepsilon _{k}=0$ and $\varepsilon
_{k}=\mu $ we have $\xi _{k}=-\mu $ and $\xi _{k}=0,$ respectively, Eq. (\ref%
{dam}) becomes

\begin{equation}
W_{P}\int_{-\mu }^{0}\frac{\left( 1-n_{\xi a}-n_{\xi b}\right) }{\sqrt{
\Delta ^{2}+\xi ^{2}}}\,\Omega \left( \xi \right) d\xi =1,  \label{dem}
\end{equation}
$\Omega \left( \xi \right) $ standing for the density of states per unit
cell. Owing to the expected smallness of $\Delta $ with respect to $\mu ,$
the main contribution to the integrand arises for $\xi \simeq 0$ so that we
can put $\Omega \left( \xi \right) \simeq \Omega _{F}$. For $\xi =0,$ we
have indeed $\varepsilon _{k}=\mu =\varepsilon _{F}$ since the chemical
potential, apart from a small correction due to temperature, coincides with
the electron kinetic energy at the Fermi level (\footnote{%
) See for instance $\left[ 33\right] $ Ch. III.}). Consequently, at $T=0$ K
where $n_{\xi a}=n_{\xi b}=0,$ Eq. (\ref{dem}) yields

\begin{equation}
W_{P}\Omega _{F}\int_{-\mu }^{0}\frac{d\xi }{\sqrt{\Delta _{0}^{2}+\xi ^{2}}}%
=W_{P}\Omega _{F}\log \frac{\Delta _{0}}{\sqrt{\Delta _{0}^{2}+\mu ^{2}}-\mu 
}\simeq W_{P}\Omega _{F}\log \frac{2\mu }{\Delta _{0}}=1  \label{dim}
\end{equation}%
which leads to (\footnote{%
) In this equation $\Delta _{0}$ is a constant quantity. But, if in
normalizing the TB functions (\ref{pat}), overlap integrals of copper ion
orbitals are not disregarded, $\Delta _{0}$ is substituted by a quantity $%
\Delta \left( \theta \right) $ depending on the angle between the electron
wave vector $\mathbf{k}$ and the cell $a$-axis. This accounts for the $d$%
-symmetry of the order parameter. In Ref. $\left[ 3\right] $ this matter has
been thoroughly discussed.})

\begin{equation}
\Delta _{0}=2\mu \exp \left( -1/W_{P}\Omega _{F}\right) .  \label{don}
\end{equation}%
This equation differs from the corresponding one of the BCS theory only in
the substitution of the Debye energy $\hbar \omega _{D}$ with twice the
chemical potential $\mu $. It follows from Eq. (\ref{don}) that
superconduction is ruled by three parameters, that is, pair energy $W_{P}$,
density of states $\Omega _{F}$ at the Fermi level and chemical potential $%
\mu .$ Taking into account that $\mu =\varepsilon _{F},$ a connexion between 
$\mu $ and $\Omega _{F}$ is expected$,$ depending on the actual electron
energy spectrum (\footnote{%
) For a tridimensional Fermi gas of $N$\ electrons in a volume $V$, we have: 
$N\,\Omega _{F}=\left( V/2\pi ^{2}\right) \left( 2m/%
h{\hskip-.2em}\llap{\protect\rule[1.1ex]{.325em}{.1ex}}{\hskip.2em}%
^{2}\right) ^{3/2}\sqrt{\varepsilon _{F}},$ where: $\varepsilon _{F}=\left( 
h{\hskip-.2em}\llap{\protect\rule[1.1ex]{.325em}{.1ex}}{\hskip.2em}%
^{2}/2m\right) \left( 3\pi ^{2}N/V\right) ^{2/3}.$ This, by letting $\mu
=\varepsilon _{F},$ leads to: $\mu \,\Omega _{F}=1.5\,$(see for instance $%
\left[ 33\right] $ Ch.\ III).}). In Appendix B, utilizing special
assumptions for the copper ions orbitals, this connexion is found to be: $%
\mu \Omega _{F}=1.28,$ which allows Eq. (\ref{don}) to be rewritten as

\begin{equation}
\Delta _{0}=2\mu \exp \left( -\,0.78\frac{\mu }{W_{P}}\right) .  \label{jiu}
\end{equation}

For $T>0$, substituting

\begin{equation}
n_{\xi a}=n_{\xi b}=1/\left[ \exp \left( \sqrt{\Delta ^{2}+\xi ^{2}}
/kT\right) +1\right]  \label{dum}
\end{equation}
into Eq. (\ref{dem}), we obtain as in the BCS theory

\begin{equation}
\log \frac{\gamma \,\Delta _{0}}{\pi k_{B}T}=\frac{7\zeta \left( 3\right) }{
8\pi ^{2}}\left( \frac{\Delta }{k_{B}T}\right) ^{2},  \label{fam}
\end{equation}
where $k_{B}$ is the Boltzmann constant, log $\gamma $ $=0.577$ the Euler
constant and $\zeta \left( 3\right) =1.202$ the Riemann Zeta-function. By
putting $\Delta =0$ in this equation, the usual relationship between energy
gap and critical temperature is found

\begin{equation}
2\Delta _{0}=\frac{2\pi }{\gamma }\,k_{B}T_{c}=3.52k_{B}T_{c}.  \label{fem}
\end{equation}%
Likewise, by utilizing the Hamiltonian (\ref{cam}) and Eqs. (\ref{com}) and (%
\ref{cum}), the quasi-particle energy spectrum is found to be the same as
that of the BCS theory, that is, 
\begin{equation}
w_{k}=\sqrt{\Delta ^{2}+\left( \varepsilon _{k}-\varepsilon _{F}\right) ^{2}}%
.  \label{fim}
\end{equation}%
The actual magnitude of the energy gap is $2\Delta _{0}$ since
quasi-particles appear in pairs as occurs in BCS\ theory. It is to pointed
out that Eq. (\ref{fem}) has been successfully tested for various
unconventional superconductors. An extensive tabulation of data concerned is
given in $\left[ 30\right] $ Ch. 6$.\medskip $

{\Large 6.} {\Large Experimental evidence in favour of the interacting-layer
mechanism }

We will now briefly examine some experimental results which substantiate the
interacting-layer mechanism. They are reported here in order of increasing
significance.

\textit{1) Coherence length} - A first clue about the interaction between
the superconducting layers is offered by measurements of the Hall effect on
YBa$_{2}$Cu$_{3}$O$_{7}$ single crystals. It was found that the
Ginzburg-Landau coherence length along the c-axis is 1.5\ $\overset{\text{o}}%
{\text{A}}.\;\func{Si}$nce the spacing of superconducting layers is near to
c-axis lattice parameter 11.68 $\overset{\text{o}}{\text{A}},$ the
conclusion was drawn that ''\textit{the two copper oxide planes which are
spaced 3.2\ }$\overset{\text{o}}{\text{A}},$\textit{\ are tightly coupled
and act as a single superconducting layer}'' $\left[ 34\right] .$

\textit{2) The effect of internal pressure}. - More direct evidence comes
out from the effect of pressure on critical temperature.\ It is common
knowledge that in cuprates T$_{\text{c}}$ considerably increases when
samples are submitted to hydrostatic pressures on the order of few GPa. Eqs.
(\ref{sec}) and (\ref{suc}), which relate pairing energy to the layer
separation $\mathbf{\lambda ,}$ explain this effect. Indeed, hydrostatic
pressure lessens separation $\mathbf{\lambda }$ thus increasing $W_{P}$ and,
consequently, critical temperature. But hydrostatic pressure lessens the
distances between copper ions in direction both parallel and orthogonal to
CuO$_{2}$ layers.\ The effect of hydrostatic pressure, therefore, is
unsuitable in distinguishing interactions inside each layer from those
between contiguous layers so that no evidence in favour of the
interacting-layer mechanism is obtained. The conclusion, however, is
different if the so-called ''internal'' or ''chemical'' pressure is
considered. Indeed, substitution of some ions with others of smaller radius
originates a decrease in the cell size which is commonly regarded as the
effect of an internal pressure. In this connection, let us quote, the
following sentence by P.\ Chu highlighted in a note by K.A.\ Muller $\left[
35\right] :$ ''\textit{Therefore, Paul Chu thought, O.K., instead of
applying pressure, I rather use a rare-earth ion, namely the yttrium which
is smaller than lanthanum, and thus get a higher T}$_{\text{c}}$\textit{\
owing to the induced internal pressure} ''. Actually, the La$^{+3}$ ionic
radius is 1.15\ $\overset{\text{o}}{\text{ A}}$ while that of Y$^{+3}$ is
0.93 $\overset{\text{o}}{\text{A}}$.\ This is tantamount to saying that in
YBCO T$_{\text{c}}$ increases just when the contiguous CuO$_{2}$ layers
approach each other leaving unchanged copper ion distances parallel to the
layers (\footnote{%
) A clear evidence of the effect of internal pressure is offered by some
thallium-based cuprates in which barium-strontium substitution slightly
lessens the cell size along c-axis so originating T$_{\text{c}}$
enhancements as large as tens of K $\left[ 36\right] .$})$.$

\textit{3) The effects of yttrium-praseodymium substitution and of reduction
in }YBa$_{2}$Cu$_{3}$O$_{7}$.\ - Concerning the effect of ion substitutions
in YBa$_{2}$Cu$_{3}$O$_{7},$ a surprising feature is the lack of
superconductivity of the PrBa$_{2}$Cu$_{3}$O$_{7}$ compound $\left[ 37,\,38%
\right] .$ This fact cannot be ascribed to the praseodymium ionic radius
which is not significantly different from the yttrium radius.\ To explain
this peculiar result it should be taken into account that praseodymium
originates both trivalent and tetravalent ions. Actually, magnetic
susceptibility measurements have indicated that, in the compound dealt with,
praseodymium is tetravalent $\left[ 39,\,40\right] .$ This means that one
electron is released from praseodymium and transferred to the neighbouring
copper ions on the CuO$_{2}$ layers so that each cell contains one divalent
and one monovalent copper rather than two divalent coppers as in the yttrium
compound. In the PrBa$_{2}$Cu$_{3}$O$_{7}$ compound praseodymium acts as an
electron donor. Since monovalent copper shows the $\left[ \text{Ar}\right] $%
3d$^{10}$ configuration lacking unpaired electrons, the superconducting
pairs can no longer be originated.\ Consequently, superconductivity is shut
out by the interacting-layer mechanism, just as expected.

A very interesting result concerns the effect of reduction on YBa$_{2}$Cu$%
_{3}$O$_{7}$ critical temperature.\ Samples of stoichiometry YBa$_{2}$Cu$%
_{3} $O$_{7-x}$ show a decreasing T$_{\text{c}}$ for increasing $x$.
Actually, the T$_{\text{c}}$ versus $x$ plot is characterized by two
plateaux, the first at 92 K, for $x$ less than 0.2, followed by a step
decrease and by a second plateau at about 60\ K, for $x$ near to 0.4.\
Measurements of distances of copper from nearby ions have shown that the
effective valence $(\footnote{%
) For the meaning of the ''effective valence'' parameter, otherwise referred
to as the ''bond valence sum'', see \ $\left[ 41,\,42,.43\right] .\;$ It can
be identified, in practice, with the copper average valence.})$ of copper on
CuO$_{2}$ layers is characterized by a parallel behaviour.\ Indeed, the plot
of copper effective valence versus $x$ shows just two plateaux for the same
values of $x$ separated by a step decrease of 0.03 e/Cu effective charges $%
\left[ 44\right] $. This behaviour, like that originated by the
yttrium-praseodimium substitution, depends on reduction of divalent copper
on the CuO$_{2}$ layers. When oxygen is removed, a number of electrons is
left into the lattice.\ Along the plateaux, only trivalent copper on the
cell basal planes is reduced thus leaving T$_{\text{c}}$ unaffected.\ The T$%
_{\text{c}}$ decrease for $0.2<x<0.4$ corresponds to the decrease of
effective valence of CuO$_{2}$ layer copper. In Reference $\left[ 5\right] $
a thorough thermodinamic description of this effect is given.

\textit{4) The monolayered }Tl$_{2}$Ba$_{2}$CuO$_{6}$\textit{\ compound}. -
Several multilayered thallium-based superconductors are known. The compond
mentioned here represents a special case since it is characterized by a
single CuO$_{2}$ layer interposed between two BaO and two TlO layers on the
outside of the BaO layers (see $\left[ 30\right] $ Ch.\ 7). With this
structure, the CuO$_{2}$ layers are well separated so that the
interacting-layer mechanism cannot be active. In spite of this, this
compound superconducts at 85\ K\ $\left[ 13\right] $.\ This fact may seem to
represent strong evidence against the mechanism we consider.\ In reality,
the situation is quite opposite.\ In fact, only reduced samples of
stoichiometry Tl$_{2}$Ba$_{2}$CuO$_{6-\delta }$ superconduct.\ Experiments
have shown that in fully oxidized samples superconductivity is destroyed
just as expected on the ground of the mechanism dealt with $\left[ 45\right] 
$. Reduction introduces in the lattice unpaired electrons, as occurs in the
perovskites with fractional stoichiometry listed in item \textit{A)} of
Section 2. As it will be shown in the next Section, these electrons
originate additional layers of unpaired electrons with which electrons in
the CuO$_{2}$ layers interact so allowing for superconductivity. On this
ground, the sudden onset of superconductivity observed as soon as the oxygen
content is reduced is not surprising.\ Indeed, even a minimum quantity of
superconducting material embedded in an inert matrix is sufficient to set
the sample resistance to zero.\ 

By keeping the previous arguments in mind, the queer result that reduction
sometimes causes and other times hinders superconductivity is explained. In
Ref.\ $\left[ 4\right] ,$ other evidences in favour of the interacting-layer
mechanism are discussed.\medskip

{\Large 7. Superconductivity of mixed stoichiometry}

{\Large perovskites }

We extend our analysis to superconductivity of some mixed stoichiometry
perovskites. As pointed out in Section 2, these materials are to be regarded
as reduced compounds. Reduction decreases the actual cation valence, thus
introducing unpaired electrons into the lattice. In the mechanism we
consider, unpaired electrons are indeed the basic ingredient for
superconductivity. However, the question is to be settled of the lattice
structure which allows the interacting layer mechanism to operate. While
cuprates show a quite tidy layered structure, in the mixed stoichiometry
compounds oxygen lacunae or substitutional ions are placed at random.
Despite this, for mere statistical reasons it can be expected that the
lattice contains some domains in which ions are layered in the right order
to allow superconductivity. We point out, in this connection, that the
formation of Lewis pairs is allowed even when lattice vectors $\mathbf{v}%
_{q} $ and $\mathbf{u}_{p}$ are not ordered on plane surfaces, as occurs in
cuprates. Even irregularly bent surfaces, such as those that are likely
found in mixed stoichiometry compounds, are suitable, provided that a number
of orbitals with unpaired electrons exist such that Eq. (\ref{Aa}) can be
applied. Therefore, remembering that even a minimum quantity of
superconducting phase sets sample resistance to zero, the superconductivity
of mixed stoichiometry perovskites can reasonably be explained.

To understand how this can occur, let us focus attention on the previously
cited 30 K superconductor Ba$_{0.6}$K$_{0.4}$BiO$_{3}$. The most
conservative assumption is that the unpaired electrons introduced by the
potassium-barium substitution lie right on the barium, so that monovalent Ba$%
^{+1}$ ions substitute K$^{+1}$ ions. In this way, in fact, the lattice
Madelung energy is kept unchanged at its former value. Evidence in favour of
this assumption is offered by the fact that superconductivity was detected
in the Sr$_{0.5}$K$_{0.5}$BiO$_{3}$ and Sr$_{0.5}$Rb$_{0.5}$ BiO$_{3}$
compounds at 12 K and 13 K, respectively (see Table 1). This large $T_{c}$
decrease is to be ascribed to the strontium ionic radius which is smaller
than the barium radius. Since the alkaly-earth ions are placed at the centre
of the perovskitic cells, the smaller Sr$^{+1}$ ion radius reduces the
orbital overlap and thus pairing energy $W_{P}$. In Table 2, using the edge,
face, centre $\left[ \text{E F C}\right] $ notation $\left[ 30,\,46\right] $%
, the probable layering scheme of Ba$_{0.6}$K$_{0.4}$BiO$_{3}$ is shown
together with those of the previously mentioned BaPb$_{0.7}$Bi$_{0.3}$O$_{3}$
and SrTiO$_{3-\delta }$ compounds. The ion arrangement which originates
superconductivity is the same in all compounds. In particular, in all
compounds an empty C-position separates the alkaly-earth ions, thus allowing
formation of bonds as occurs in bilayered cuprates.

\begin{equation*}
\begin{tabular}{ccc}
&  & $\left[ \text{BiO}_{2}-\right] $ \\ 
$\left[ \text{O}-\text{Sr}\right] $ & $\left[ \text{O}-\text{Ba}\right] $ & $%
\left[ \text{O}-\text{K}\right] $ \\ 
$\left[ \text{TiO}_{2}-\right] $ & $\left[ \text{PbO}_{2}-\right] $ & $\left[
\text{BiO}_{2}-\right] $ \\ 
$\left\{ 
\begin{array}{c}
\left[ --\text{\textbf{Sr}}^{\text{+1}}\right] \\ 
\left[ \text{TiO}_{2}-\right] \\ 
\left[ --\text{\textbf{Sr}}^{\text{+1}}\right]%
\end{array}%
\right\} $ & $\,\,\,\,\,\,\left\{ 
\begin{array}{c}
\left[ \text{O}-\text{\textbf{Ba}}^{\text{+1}}\right] \\ 
\left[ \text{BiO}_{2}-\right] \\ 
\left[ \text{O}-\text{\textbf{Ba}}^{\text{+1}}\right]%
\end{array}%
\right\} \,\,\,\,\,$\thinspace & $\left\{ 
\begin{array}{c}
\left[ \text{O}-\text{\textbf{Ba}}^{\text{+1}}\right] \\ 
\left[ \text{BiO}_{2}-\right] \\ 
\left[ \text{O}-\text{\textbf{Ba}}^{\text{+1}}\right]%
\end{array}%
\right\} $ \\ 
$\left[ \text{TiO}_{2}-\right] $ & $\left[ \text{BiO}_{2}-\right] $ & $\left[
\text{BiO}_{2}-\right] $ \\ 
$\left[ \text{O}-\text{Sr}\right] $ & $\left[ \text{O}-\text{Ba}\right] $ & $%
\left[ \text{O}-\text{K}\right] $ \\ 
SrTiO$_{3-\delta }$ & $\left[ \text{PbO}_{2}-\right] $ & $\left[ \text{BiO}%
_{2}-\right] $ \\ 
T$_{\text{c}}=0.3$ K & BaPb$_{0.7}$Bi$_{0.3}$O$_{3}$ & $\left[ \text{O}-%
\text{K}\right] $ \\ 
& T$_{\text{c}}=10$ K & Ba$_{0.6}$K$_{0.4}$BiO$_{3}$ \\ 
&  & T$_{\text{c}}=30$ K%
\end{tabular}%
\end{equation*}

$\,$%
\begin{equation*}
\,%
\begin{tabular}{cc}
& $\left[ -\text{O}_{2}\text{Cu}\right] $ \\ 
& $\left[ \text{Ba}-\text{O}\right] $ \\ 
$\left[ -\text{O}_{2}\text{Cu}\right] $ & $\left[ \text{O}-\text{Tl}\right] $
\\ 
$\left[ \text{La}-\text{O}\right] $ & $\left[ \text{Tl}-\text{O}\right] $ \\ 
$\left\{ 
\begin{array}{c}
\left[ \text{O}-\text{\textbf{Sr}}^{\text{+3}}\right] \\ 
\left[ \text{\textbf{Cu}O}_{2}-\right]%
\end{array}%
\right\} \,\,$ & $\,\,\,\,\,\left\{ 
\begin{array}{c}
\left[ --\text{\textbf{Ba}}^{\text{+1}}\right] \\ 
\left[ \text{\textbf{Cu}O}_{2}-\right]%
\end{array}%
\right\} $ \\ 
$\left[ \text{O}-\text{La}\right] $ & $\left[ \text{O}-\text{Ba}\right] $ \\ 
$\left[ \text{La}-\text{O}\right] $ & $\left[ \text{Tl}-\text{O}\right] $ \\ 
$\left[ -\text{O}_{2}\text{Cu}\right] $ & $\left[ \text{O}-\text{Tl}\right] $
\\ 
La$_{1.85}$Sr$_{0.15}$CuO$_{4}$ & $\left[ \text{Ba}-\text{O}\right] $ \\ 
T$_{\text{c}}=35$ K & $\left[ -\text{O}_{2}\text{Cu}\right] $ \\ 
& Tl$_{2}$Ba$_{2}$CuO$_{6-\delta }$ \\ 
& T$_{\text{c}}=85$ K%
\end{tabular}%
\end{equation*}

Table 2: Probable layering schemes in mixed stoichiometry perovskites.
Braces enclose the layers which activate superconductivity. Ions with
unpaired electrons are represented by bold type.\medskip

Probable layering schemes of La$_{1.85}$Sr$_{0.15}$CuO$_{4}$ and Tl$_{2}$Ba$%
_{2}$CuO$_{6-\delta }$ compounds are also shown in Table 2. These cuprates
are of special interest since both are characterized by isolated CuO$_{2}$
layers. In the oxidized La$_{1.85}$Sr$_{0.15}$CuO$_{4}$ compound, trivalent
Sr$^{+3}$ ions are included showing unpaired electrons in the krypton shell
and forming bonds with copper. The situation, however, is quite different
from that of bilayered cuprates. In fact, while copper is placed in
E-position, strontium lies in the C-position of the overhanging layer.
Consequently, each copper is allowed to form bonds with four strontium ions
so that the symmetry between the interacting layers peculiar to bilayered
cuprates no longer exists.\ A possible equivalent interpretation assumes
that divalent strontium causes an equal number of lanthanum ions to be
oxidized to valence four, thus showing unpaired electrons in the xenon
shell. The monolayered reduced compound Tl$_{2}$Ba$_{2}$CuO$_{6-\delta }$
shows a similar situation.\ In this compound, monovalent Ba$^{+1}$ ions with
unpaired 6s electrons are present forming bonds with 3d-electrons of
divalent copper. Both the compounds dealth with are indeed characterized by
staggered overlaps of the interacting layers (\footnote{%
) A state of affairs of this kind has been already considered in Ref. $\left[
3\right] $.}). This would require some modifications to calculations of
Section 4, leaving however the essential results unchanged. In reality, the
matters presented in this Section are based in part on conjectures owing to
the lack of data on the actual placing of the unpaired electrons.\
Notwithstanding this, in our opinion, the reliability of the interacting
layer mechanism is reasonably proved.\medskip

{\Large 8. Discussion and conclusions}

According to the electron pairing mechanism we propose, superconductivity in
cuprates requires the presence of two neighbouring CuO$_{2}$ layers. In
compounds like YBa$_{2}$Cu$_{3}$O$_{7},$ each couple of layers constitutes
an independent superconductor. Contrary to this point of view, evidence has
been claimed for nonexistence of superconductivity in an isolated CuO$_{2}$
bilayer. Organic chains (Py-C$_{\text{n}}$H$_{2\text{n}+1}$)$_{2}$HgI$_{4}$
(2\TEXTsymbol{<}n\TEXTsymbol{<}12) were intercalated in the bilayered Bi$%
_{2} $Sr$_{2}$CaCu$_{2}$O$_{8}$ compound thus drastically increasing the
distance between consecutive bilayers $\left[ 47\right] $. In this way, a
complete disappearance of superconductivity was observed. This result was
considered as a proof that superconductivity depends on a three-dimensional
linkage between the couples of neighbouring CuO$_{2}$ layers.\ In opposition
to this conclusion, we point out that pyridine is an efficient electron
donor (see for instance $\left[ 48\right] $).\ Consequently, the observed
disappearance of superconductivity is an expected donor-effect similar to
that of yttrium-praseodimium substitution in YBa$_{2}$Cu$_{3}$O$_{7}$
discussed in item \textit{3)} of Section 6.

The question of the actual number of layers required for originating
superconductivity in cuprates is certainly of primary importance.\ The idea
that interlayer coupling plays a role in superconductivity dates from 1987
when \ Z. Te\v{s}anovi\'{c} proposed a mechanism that involves Coulomb
interaction from the band at the Fermi surface to some of the fully occupied
or empty bands away from the Fermi level $\left[ 49\right] $. Another
momentous question is the basic interaction which allows the formation of
electron pairs.\ In alternative to the phonon coupling peculiar to the BCS
theory, in 1997 we proposed the inter-layer HL-type two-electron exchange $%
\left[ 2\right] ,$ while T.M.\ Mishonov et al. proposed an inter-atomic
two-electron exchange $\left[ 50\right] $. In contrast with this previous
proposals, these authors recently have advanced the intra-atomic exchange of
two electrons between 4s and 3d$_{x^{2}-y^{2}}$ states as the origin of high
T$_{\text{c}}$ superconduction in cuprates $\left[ \,51\right] $. In our
opinion, identification of exchange interactions as the very cause of
superconductivity represents a major progress in this field of studies. But,
for a full understanding of the phenomenon, the question remains to be
settled if an unique mechanism or different mechanisms are active in the
different kinds of unconventional superconductors.

Leaving aside the theoretical issues, the most pressing thing appears to be
the discovery of new superconductors of higher T$_{\text{c}}$ and,
hopefully, of a 300 K superconductor. In reality, even a minor increase of
pairing energy $W_{P}$ may originate a large increase of T$_{\text{c}}$,
owing to the exponential dependence of $\Delta _{0}\ $on$\ W_{P}\ $given in
Eq. (\ref{jiu}). A sound argument in favour of this expectation is offered
by the detection of a sharp superconductive transition at 235 K due to
traces of an unidentifyed phase fortuitously mixed to a HgBa$_{2}$Ca$_{2}$Cu$%
_{3}$O$_{8}$ sample $\left[ 52\right] .$ The approach we advise for
improving critical temperature is based on the fact that all cuprates with
divalent calcium sandwiched between the neighbouring CuO$_{2}$ layers show
values of T$_{\text{c}}$ higher than 92 K, the YBa$_{2}$Cu$_{3}$O$_{7}$
critical temperature, in which trivalent yttrium is sandwiched between the
layers. This fact can be explained by considering that the charge of
unpaired electrons on copper ions is a little shifted towards the sandwiched
positive ions, thus reducing overlap of unpaired electrons and,
consequently, $W_{P}.\;$Obviously, this effect is expected to be less
harmful with divalent calcium than with trivalent yttrium.$\;$This induces
us to consider compounds of stoichiometry M$^{+1}$M$_{2}^{+3}$Cu$_{3}$O$_{7}$
derived from YBa$_{2}$Cu$_{3}$O$_{7}$ by substituting \thinspace Y$^{+3}$
with monovalent \thinspace M$^{+1}$ ions and Ba$^{+2}$ with trivalent M$%
^{+3} $ ions (M$^{+1}=\,$Li$,$ Na, K; M$^{+3}=$ Y, La).\ This substitution
leaves the cell neutrality unchanged. Using the $\left[ \text{E, F, C}\right]
$ notation $\left[ 8,\,9\right] $, the substitution spoken of is shown in
Table 3. Along the same line of reasoning, in the HgBa$_{2}$Ca$_{2}$Cu$_{3}$O%
$_{8}$ compound we can consider the substitution of Ca$^{+2}$ ions with M$%
^{+1}$ ions and Ba$^{+2}$ ions with M$^{+3}$ ions yielding the HgM$_{2}^{+3}$%
M$_{2}^{+1}$Cu$_{3}$O$_{8}$ compound.\ Also the thallium based compounds Tl$%
_{2}$Ba$_{2}$CaCu$_{2}$O$_{8}$ and Tl$_{2}$Ba$_{2}$Ca$_{2}$Cu$_{3}$O$_{10}$
are in principle suitable for substitution of calcium with monovalent ions.
Of course, the possibility of obtaining these substituted compounds is a
mere conjecture which should be confirmed by experiments.\ 

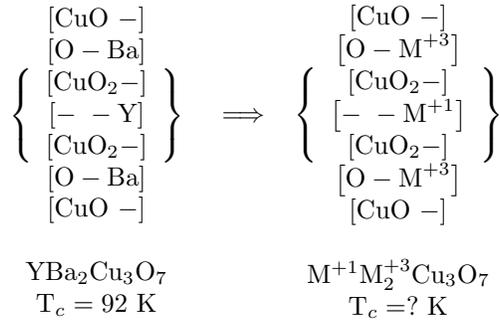
\begin{equation*}
\end{equation*}

\begin{equation*}
\begin{tabular}{c}
$\left[ \text{CuO\thinspace \thinspace }-\right] $ \\ 
$\left[ \text{O}-\text{Ba}\right] $ \\ 
$\left\{ 
\begin{array}{c}
\left[ \text{CuO}_{2}-\right] \\ 
\left[ -\,\,-\text{Y}\right] \\ 
\left[ \text{CuO}_{2}-\right]%
\end{array}%
\right\} $ \\ 
$\left[ \text{O}-\text{Ba}\right] $ \\ 
$\left[ \text{CuO\thinspace \thinspace }-\right] $ \\ 
\\ 
YBa$_{2}$Cu$_{3}$O$_{7}$ \\ 
T$_{c}=92$ K%
\end{tabular}%
\ 
\begin{array}{c}
\Longrightarrow \\ 
\overset{}{} \\ 
\overset{}{} \\ 
\overset{}{}%
\end{array}%
\begin{tabular}{c}
$\left[ \text{CuO\thinspace \thinspace }-\right] $ \\ 
$\left[ \text{O}-\text{M}^{+3}\right] $ \\ 
$\left\{ 
\begin{array}{c}
\left[ \text{CuO}_{2}-\right] \\ 
\left[ -\,\,-\text{M}^{+1}\right] \\ 
\left[ \text{CuO}_{2}-\right]%
\end{array}%
\right\} $ \\ 
$\left[ \text{O}-\text{M}^{+3}\right] $ \\ 
$\left[ \text{CuO\thinspace \thinspace }-\right] $ \\ 
\\ 
M$^{+1}$M$_{2}^{+3}$Cu$_{3}$O$_{7}$ \\ 
T$_{c}=?$ K%
\end{tabular}%
\end{equation*}

\begin{eqnarray*}
&&\quad \ \ 
\begin{tabular}{c}
$\left[ \text{Hg}-\,-\right] $ \\ 
$\left[ \text{O}-\text{Ba}\right] $ \\ 
$\left\{ 
\begin{array}{c}
\left[ \text{CuO}_{2}-\right] \\ 
\left[ --\text{Ca}\right] \\ 
\left[ \text{CuO}_{2}-\right] \\ 
\left[ --\text{Ca}\right] \\ 
\left[ \text{CuO}_{2}-\right]%
\end{array}%
\right\} $ \\ 
$\left[ \text{O}-\text{Ba}\right] $ \\ 
$\left[ \text{Hg}-\,-\right] $ \\ 
\\ 
HgBa$_{2}$Ca$_{2}$Cu$_{3}$O$_{8}$ \\ 
T$_{c}=133$ K%
\end{tabular}%
\begin{array}{c}
\Longrightarrow \\ 
\overset{}{} \\ 
\overset{}{} \\ 
\overset{}{}%
\end{array}%
\begin{tabular}{c}
$\left[ \text{Hg}-\,-\right] $ \\ 
$\left[ \text{O}-\text{M}^{+3}\right] $ \\ 
$\left\{ 
\begin{array}{c}
\left[ \text{CuO}_{2}-\right] \\ 
\left[ -\,\,-\text{M}^{+1}\right] \\ 
\left[ \text{CuO}_{2}-\right] \\ 
\left[ -\,\,-\text{M}^{+1}\right] \\ 
\left[ \text{CuO}_{2}-\right]%
\end{array}%
\right\} $ \\ 
$\left[ \text{O}-\text{M}^{+3}\right] $ \\ 
$\left[ \text{Hg}-\,-\right] $ \\ 
\\ 
HgM$_{2}^{+3}$M$_{2}^{+1}$Cu$_{3}$O$_{8}$ \\ 
T$_{c}=?$ K%
\end{tabular}
\\
&&
\end{eqnarray*}%
Table 3: \ Layering schemes of YBa$_{2}$Cu$_{3}$O$_{7}$ and HgBa$_{2}$Ca$%
_{2} $Cu$_{3}$O$_{8}$ cuprates and their respective modified counterparts M$%
^{+1}$M$_{2}^{+3}$Cu$_{3}$O$_{7}$ and HgM$_{2}^{+3}$M$_{2}^{+1}$Cu$_{3}$O$%
_{8}$. Braces enclose the layers which activate superconductivity.\bigskip

\medskip

\bigskip {\Large Appendix \ A - Some remarks about the Belyaev Hamiltonian
for pairing correlations in fermion systems}

S.T. Belyaev during a stay at the Institute for Theoretical Physics of the
University of Copenhagen wrote a paper entitled ''Effect of pairing
correlations on \ nuclear properties''. We report hereinafter Subsection 1
(pag. 7) of this paper (\footnote{%
) Published in Matematisk-fysiske Meddelelser (31, no. 11, 1959) a journal
issued by the Royal Danish Academy of Sciences and Letters.}) in which the
Hamiltonian for the system of interacting particles is given.

\begin{center}
$\circ \circ \circ \circ \circ \circ \circ \circ \circ \circ \circ \circ
\circ \circ \circ \circ \circ \circ \circ \circ \circ \circ \circ \circ
\circ \circ \circ \circ \bigskip $

{\large 1.\ Hamiltonian\ }
\end{center}

We consider a system of nucleons which are moving in a certain axially
symmetric self-consistent well. (For simplicity, we do not distinguish
between neutrons and protons). As basic functions of the second quantization
representation we choose the wave functions of a nucleon in this well.
States which differ only in the sign of the projections of angular momentum
along the symmetry axis are degenerate.\ We call such states
\textquotedblright conjugate\textquotedblright\ states and mark them with
the index \ $k\sigma =\left( k+;\text{ }k-\right) ^{\ast }.$

The wave functions of the conjugate states are assumed to transform into
each other by complex conjugation and exchange of the spinor components$%
^{\ast \ast }.$

Let us introduce the Fermi operators $a_{k\sigma }^{+};$ $a_{k\sigma }^{{}}$%
which create and destroy a particle in the state $k\sigma .$ The Hamiltonian
for the system of interacting particles is then%
\begin{equation*}
H^{\prime }=\sum_{k}\varepsilon _{k}\left(
a_{k+}^{+}a_{k+}^{{}}+a_{k-}^{+}a_{k-}^{{}}\right)
\end{equation*}%
\begin{equation*}
-\frac{1}{2}\sum_{\left( k,\text{ }\sigma \right) }\left\langle k_{1}\sigma
_{1}k_{2}\sigma _{2}\right| G\left| k_{2}^{\prime }\sigma _{2}^{\prime
}k_{1}^{\prime }\sigma _{1}^{\prime }\right\rangle a_{k_{1}\sigma
_{1}}^{+}a_{k_{2}\sigma _{2}}^{+}a_{k_{2}^{\prime }\sigma _{2}^{\prime
}}^{{}}a_{k_{1}^{\prime }\sigma _{1}^{\prime }}^{{}},\qquad \left( 1\right)
\end{equation*}%
where $\varepsilon _{k}$ is the single-particle energy in $k$-th state. (The
sign of G is chosen to be positive for an attractive interaction). The
Hamiltonian (1) describes a system with a fixed number of particles $N$.
Therefore, in a perturbation treatment in which $H^{\prime }$ is split into
two parts, each of these parts must commute with $N$. The problem is
essentially simplified if we make a transition from the system with fixed $N$
(''$N$-system'') to one with a fixed value of the chemical potential $%
\lambda $ (''$\lambda $-system''), which is described by the Hamiltonian%
\begin{equation*}
H=H^{\prime }-\lambda N\text{.}\qquad \qquad \qquad \qquad \left( 2\right)
\end{equation*}%
The choice of $\ \lambda $ determines only the average value of $N$ in the $%
\lambda $-system. Therefore, the solution which corresponds to the
Hamiltonian, (2), will describe only average properties of nuclei and does
not pretend to describe the individual nuclear properties for which one
needs a fixed value of $N$. As will be shown later, the uncertainty in the
value of $N$\ is small. In practice, the averaging is done only over a few
neighbouring nuclei, either all even or all odd.

\begin{quote}
\textit{Footnotes}

*) In fact, even symmetry of reflection in a plane is enough for the
definition of the conjugated states. We speak of axial symmetry only for
definiteness.

**) If $\ \psi _{+}=\binom{\psi _{1}^{{}}}{\psi _{2}^{{}}}$, then $\psi _{-}=%
\binom{\psi _{2}^{\ast }}{-\psi _{1}^{\ast }}$. The transformation $\psi
_{+}\rightarrow \psi _{-}$ is equivalent to the time reversal T.
\end{quote}

\begin{center}
$\circ \circ \circ \circ \circ \circ \circ \circ \circ \circ \circ \circ
\circ \circ \circ \circ \circ \circ \circ \circ \circ \circ \circ \circ
\circ \circ \circ \circ $

\bigskip
\end{center}

Let\ the Hamiltonian for pair formation, given in Eq.(1) of the previous
Belyaev's paper, be written in the form

\begin{equation}
\widehat{H}_{pair}=-\frac{1}{2}\sum_{\left( k,\text{ }\sigma \right)
}\left\langle k_{1}\sigma _{1}k_{2}\sigma _{2}\right| G\left| k_{2}^{\prime
}\sigma _{2}^{\prime }k_{1}^{\prime }\sigma _{1}^{\prime }\right\rangle 
\widehat{\alpha }_{k_{1}\sigma _{1}}^{+}\widehat{\alpha }_{k_{2}\sigma
_{2}}^{+}\widehat{\alpha }_{k_{2}^{\prime }\sigma _{2}^{\prime }}^{{}}%
\widehat{\alpha }_{k_{1}^{\prime }\sigma _{1}^{\prime }}^{{}}\text{.}
\label{xaz}
\end{equation}%
By appropriately defining indexes $k$ and $\sigma $, it assumes different
meanings. Indeed, by putting

\begin{equation}
k_{1}=k,\quad k_{2}=-k,\quad k_{1}^{\prime }=k^{\prime },\quad k_{2}^{\prime
}=-k^{\prime },\quad \quad \sigma _{1}=\sigma _{1}^{\prime }=\uparrow ,\quad
\sigma _{2}=\sigma _{2}^{\prime }=\downarrow ,  \label{xwz}
\end{equation}

\begin{equation}
\frac{1}{2}\left\langle k_{1}\sigma _{1}k_{2}\sigma _{2}\right| G\left|
k_{2}^{\prime }\sigma _{2}^{\prime }k_{1}^{\prime }\sigma _{1}^{\prime
}\right\rangle =\frac{1}{2}\left\langle k\uparrow -k\downarrow \right|
G\left| -k^{\prime }\downarrow k^{\prime }\uparrow \right\rangle
=V_{k^{\prime },\text{ }k}  \label{xcz}
\end{equation}%
and moving toward left the apex ($\prime $), $\widehat{H}_{pair}$ can be
rewritten in the quick form

\begin{equation}
\widehat{H}_{pair}=-\sum_{k,\text{ }k^{\prime }}V_{k,\text{ }k^{\prime }%
\text{ }}\widehat{\alpha }_{k^{\prime }\uparrow }^{+}\widehat{\alpha }%
_{-k^{\prime }\downarrow }^{+}\widehat{\alpha }_{-k\downarrow }^{{}}\widehat{%
\alpha }_{k\uparrow }^{{}},  \label{xdz}
\end{equation}%
where arrows\ stand for\ spin components. This is the Hamiltonian for the
Cooper pairs in the BCS theory. These pairs are characterized by opposite
electron momenta, that is, by reflection symmetry around a center. In a
different way, by allowing for the conjugated states considered in the
Belyaev's paper, we have

\begin{equation}
k_{1}=k_{2}=k,\quad k_{1}^{\prime }=k_{2}^{\prime }=k^{\prime }\quad \sigma
_{1}=\sigma _{1}^{\prime }=+,\quad \sigma _{2}=\sigma _{2}^{\prime }=-,
\label{xez}
\end{equation}%
\begin{equation}
\frac{1}{2}\left\langle k_{1}\sigma _{1}k_{2}\sigma _{2}\right| G\left|
k_{2}^{\prime }\sigma _{2}^{\prime }k_{1}^{\prime }\sigma _{1}^{\prime
}\right\rangle =\frac{1}{2}\left\langle k+,k-\right| G\left| k^{\prime
}-,k^{\prime }+\right\rangle =V_{k^{\prime },\text{ }k}  \label{xfz}
\end{equation}%
and

\begin{equation}
\widehat{H}_{pair}=-\sum_{k,\text{ }k^{\prime }}V_{k,\text{ }k^{\prime }%
\text{ }}\widehat{\alpha }_{k^{\prime }+}^{+}\widehat{\alpha }_{k^{\prime
}-}^{+}\widehat{\alpha }_{k-}^{{}}\widehat{\alpha }_{k+}^{{}}  \label{xgz}
\end{equation}%
which is the Hamiltonian for the Majorana pairs, characterized by opposite
signs of the projections of angular momenta along the symmetry axis. We
emphasize, finally, that as pointed out in footnote (*) of the Belyaev's
paper, even fermions of equal momenta which exibit symmetry of reflection in
a plane can be paired in conjugated states. The corresponding Hamiltonian
can be written as

\begin{equation}
\widehat{H}_{pair}=-\sum_{k,\text{ }k^{\prime }}V_{k,\text{ }k^{\prime }%
\text{ }}\widehat{\alpha }_{k^{\prime }R}^{+}\widehat{\alpha }_{k^{\prime
}L}^{+}\widehat{\alpha }_{kL}^{{}}\widehat{\alpha }_{kR}^{{}},  \label{xhz}
\end{equation}%
where indexes $L$ and $R$ mean left and right with respect to the reflection
plane. Symmetries of interacting fermion pairs\ are summarized in Table 4.
Pairs showing planar symmetry have been named \textquotedblright Lewis
pairs\textquotedblright . This because, as explained in Sections 4 and 5,
superconductivity in cuprates can be ascribed to pairs of electrons forming
covalent bonds characterized just by this kind of symmetry.

\begin{center}
Table 4: Symmetries of fermion pairs.%
\begin{equation*}
\begin{array}{ccc}
\text{Pairs} & \text{Conjugated states} & \text{Symmetry} \\ 
\text{Cooper} & -k,\downarrow \quad \ast \quad +k,\uparrow & \text{Central}
\\ 
\text{Majorana} & k,-\quad \ast \quad k,+ & \text{Axial} \\ 
\text{Lewis} & k,L\quad \ast \quad k,R & \text{Planar}%
\end{array}%
.
\end{equation*}

\medskip \bigskip
\end{center}

{\Large Appendix B - The electron energy spectrum}

As shown in Eq. (\ref{zzz}) and by writing briefly $W_{k}$ for $W_{a}\left( 
\mathbf{k}_{a}\right) $ and $W_{b}\left( \mathbf{k}_{b}\right) $, the energy
of electrons running on the CuO$_{2}$ layers is given by the expectation
value

\begin{equation}
W_{k}=\frac{1}{N}\left\langle \sum_{m=1}^{N}\exp \left( i\,\mathbf{k\cdot u}
_{m}\right) a\left( \mathbf{r}-\mathbf{u}_{m}\right) \right| H\left( \mathbf{%
p,r} \right) \left| \sum_{n=1}^{N}\exp \left( i\,\mathbf{k\cdot u}%
_{n}\right) a\left( \mathbf{r}-\mathbf{u}_{n}\right) \right\rangle
\label{gam}
\end{equation}
of Hamiltonian

\begin{equation}
H\left( \mathbf{p,r}\right) =\frac{\mathbf{p}^{2}}{2m}-\sum_{p=1}^{N}\frac{%
Z\,e^{2}}{\left| \mathbf{r}-\mathbf{u}_{p}\right| }+V_{lat}\left( \mathbf{r}%
\right) .  \label{gem}
\end{equation}
This energy is of course the same for $a$ or $b$ layers. For completeness
sake, in Eq. (\ref{gem}) the lattice potential $V_{lat}\left( \mathbf{r}%
\right) $ has been included due to ions in positions other than $\mathbf{u}%
_{p},$ that is, to oxygen, yttrium and barium ions neighbouring the copper
ions. Orbitals of copper ions are solutions of equation

\begin{equation}
\left[ \frac{\mathbf{p}^{2}}{2m}-\frac{Z\,e^{2}}{\left\vert \mathbf{r}-%
\mathbf{u}\right\vert }+V_{lat}\left( \mathbf{r}\right) \right] a\left( 
\mathbf{r}-\mathbf{u}\right) =E_{3d}\,a\left( \mathbf{r}-\mathbf{u}\right)
\label{gim}
\end{equation}%
$E_{3d}$ standing for the orbital energy. On the CuO$_{2}$ layers, copper
ions form a square grid with Cu$^{+2}$ ions at the square vertices and O$%
^{-2}$ ions at the middle of the square sides. The Coulomb field of O$^{-2}$
ions cuts down the electron charge density of Cu$^{+2}$ ions along the
square sides, thus increasing density along the square diagonals. So the
electron charge distribution is expected to show the four-lobe shape
peculiar to d-orbitals (\footnote{%
) As for the real form of copper orbitals in CuO$_{2}$ layers, the most
likely assumption is that they consist of a superposition of $3d_{xy}$ and $%
3d_{x^{2}-y^{2}}$ orbitals. The $3d_{x^{2}-y^{2}}$ orbitals are lined up
along the Cu-O-Cu chains, thus allowing for super-exchange interactions
between coppers mediated by oxygen ions. Consequently, only the $3d_{xy}$
contributions should be considered when dealing with pairing energy of
electrons on neighbouring CuO$_{\text{2}}$ layers (see Ref .\ $\left[ 4%
\right] $ ).}).

Cu$^{+2}$ ions placed at opposite ends of the square sides cannot interact
directly owing to the interposite oxygens. On the contrary, Cu$^{+2}$ ions
placed at opposite ends of the square diagonals are able to originate
overlap and exchange integrals of significant values. This entails that Cu$%
^{+2}$ ions placed alternatively along the square sides form two independent
but equivalent ion sets, each holding $N/2$ ions. It is therefore sufficient
to consider one of these sets. So Eq. (\ref{gam}) can be rewritten as

\begin{equation}
W_{k}=\frac{2}{N}\sum_{m,\,n=1}^{N/2}\exp \left[ i\,\mathbf{k}\cdot \left( 
\mathbf{u}_{n}-\mathbf{u}_{m}\right) \right] \int a\left( \mathbf{r}-\mathbf{%
u}_{m}\right) \,H\left( \mathbf{p,r}\right) \,a\left( \mathbf{r}-\mathbf{u}%
_{n}\right) d^{3}\mathbf{r.}  \label{gom}
\end{equation}
By taking into account that for $n\neq m$ only ions in the four neighbouring
lattice positions $\mathbf{u}_{i}$ around $\mathbf{u}_{m}$ make a
significant contribution, we have

\begin{equation*}
W_{k}=\frac{2}{N}\sum_{m=1}^{N/2}\left[ \int a\left( \mathbf{r}_{m}\right)
\,H\left( \mathbf{p,r}_{m}\right) \,a\left( \mathbf{r}_{m}\right) d^{3}%
\mathbf{r}_{m}\right. +
\end{equation*}

\begin{equation}
\left. \sum_{i=1}^{4}\exp \left( i\,\mathbf{k}\cdot \mathbf{\sigma }%
_{i}\right) \,\int a\left( \mathbf{r}_{i}+\mathbf{\sigma }_{i}\right)
\,H\left( \mathbf{p,r}_{i}\right) \,a\left( \mathbf{r}_{i}\right) d^{3}%
\mathbf{r}_{i}\right] ,  \label{gum}
\end{equation}%
where $\mathbf{r}_{m}=\mathbf{r}-\mathbf{u}_{m}\mathbf{,}$ $\mathbf{r}_{i}=%
\mathbf{r}-\mathbf{u}_{i}$ and $\mathbf{\sigma }_{i}=\mathbf{u}_{i}-\mathbf{u%
}_{m}.$ For an unlimited CuO$_{2}$ layer, the sum over $m$ is independent of
position $\mathbf{u}_{m}$ so that all terms in the sum are equal. This
allows us to write

\begin{equation*}
W_{k}=\int a\left( \mathbf{r}\right) \,H\left( \mathbf{p,r}\right) \,a\left( 
\mathbf{r}\right) d^{3}\mathbf{r}+
\end{equation*}

\begin{equation}
+\sum_{i=1}^{4}\exp \left( i\,\mathbf{k}\cdot \mathbf{\sigma }_{i}\right)
\,\int a\left( \mathbf{r}_{i}+\mathbf{\sigma }_{i}\right) \,H\left( \mathbf{%
p,r}_{i}\right) \,a\left( \mathbf{r}_{i}\right) d^{3}\mathbf{r}_{i}.
\label{lam}
\end{equation}%
On the other hand, we have from Eqs. (\ref{gim}) and (\ref{gem})

\begin{equation}
H\left( \mathbf{p,r}_{m}\right) a\left( \mathbf{r}_{m}\right) =\left(
E_{3d}-\sum_{i=1}^{4}\frac{Z\,e^{2}}{\left| \mathbf{r}_{i}\right| }\right)
a\left( \mathbf{r}_{m}\right) ,  \label{lem}
\end{equation}
in which, as for Eq. (\ref{gum}), only terms for ions in the four positions $%
\mathbf{u}_{i}$ around $\mathbf{u}_{m}$ have been included. Owing to square
symmetry of these positions around $\mathbf{u}_{m}$, we obtain

\begin{equation}
\int a\left( \mathbf{r}\right) \,H\left( \mathbf{p,r}\right) \,a\left( 
\mathbf{r} \right) d^{3}\mathbf{r=\,}E_{3d}-4E_{C},  \label{lim}
\end{equation}
where

\begin{equation}
E_{C}=\int a\left( \mathbf{r}\right) \,\frac{Z\,e^{2}}{\left| \mathbf{r}-%
\mathbf{\sigma }\right| }\,a\left( \mathbf{r}\right) d^{3}\mathbf{r}
\label{lom}
\end{equation}
means Coulomb energy of the ion in position $\mathbf{u}$ originated by the
electric field of the ion in position $\mathbf{u}+\mathbf{\sigma }.$ By
putting $\mathbf{r}_{p}=\mathbf{r}-\mathbf{u}_{p},$ we have analogously

\begin{equation*}
\int a\left( \mathbf{r}_{i}+\mathbf{\sigma }_{i}\right) \,H\left( \mathbf{p,r%
}_{i}\right) \,a\left( \mathbf{r}_{i}\right) d^{3}\mathbf{r}_{i}\mathbf{=}
\end{equation*}

\begin{equation}
=E_{3d}\int a\left( \mathbf{r}_{i}+\mathbf{\sigma }_{i}\right) \,a\left( 
\mathbf{r}_{i}\right) d^{3}\mathbf{r}_{i}-\sum_{p\left( \neq \,i\right)
=1}^{4}\int a\left( \mathbf{r}_{i}+\mathbf{\sigma }_{i}\right) \frac{Z\,e^{2}%
}{\left\vert \mathbf{r}_{p}\right\vert }\,a\left( \mathbf{r}_{i}\right) d^{3}%
\mathbf{r}_{i}.  \label{lum}
\end{equation}%
Since the product $a\left( \mathbf{r}_{i}+\mathbf{\sigma }_{i}\right)
\,a\left( \mathbf{r}_{i}\right) $ shows a significant value only midway
between the $\mathbf{u}_{m}$ and $\mathbf{u}_{i}$ positions, only the
integral for $p=m$ is to be accounted for. By omitting label $i$ and
introducing the overlap

\begin{equation}
O=\int a\left( \mathbf{r}+\mathbf{\sigma }\right) \,a\left( \mathbf{r}%
\right) d^{3}\mathbf{r}  \label{qam}
\end{equation}
and exchange

\begin{equation}
E_{ex}=\int a\left( \mathbf{r}+\mathbf{\sigma }\right) \frac{Z\,e^{2}}{%
\left| \mathbf{r}+\mathbf{\sigma }\right| }\,a\left( \mathbf{r}\right) d^{3}%
\mathbf{r}  \label{qem}
\end{equation}
integrals, we obtain

\begin{equation}
\int a\left( \mathbf{r}_{i}+\mathbf{\sigma }_{i}\right) \,H\left( \mathbf{p,r%
} _{i}\right) \,a\left( \mathbf{r}_{i}\right) d^{3}\mathbf{r}_{i}\mathbf{=}
OE_{3d}-E_{ex}.  \label{qim}
\end{equation}
In this way, substitution of Eqs. (\ref{lim}) and (\ref{qim}) into Eq. (\ref%
{lam}) leads to

\begin{equation}
W_{k}=E_{3d}-4E_{C}+\left( OE_{3d}-E_{ex}\right) \sum_{i=1}^{4}\exp \left(
i\,\mathbf{k}\cdot \mathbf{\sigma }_{i}\right) .  \label{qom}
\end{equation}
This result mimics the one obtained in the case of tridimensional lattices
in which six vectors $\mathbf{\sigma }_{i}$ have to be considered.

By assuming $x$ and $y$ axes parallel to vectors $\mathbf{\sigma }_{i},$ by
letting $W_{0}=E_{3d}-4E_{C}+4\left( OE_{3d}-E_{ex}\right) $ and $B=4\left(
OE_{3d}-E_{ex}\right) $ for ground state and band energies, respectively,
the electron kinetic energy turns out to be

\begin{equation}
\varepsilon _{k}=W_{k}-W_{0}=\frac{B}{2}\left[ 2-\cos \left( k_{x}\sigma
\right) -\cos \left( k_{y}\sigma \right) \right] ,  \label{qum}
\end{equation}
where $\sigma =\left| \mathbf{\sigma }_{i}\right| .\;$It follows that for $%
k_{x}\sigma ,\,k_{y}\sigma =\pm \pi ,$ kinetic energy attains its maximum
value $2B$. By assuming the CuO$_{2}$ plane to be a square of area $a\times
a $ with sides parallel to $x$ and $y$ axes and taking into account that $%
k_{x}=n_{x}\,\left( \pi /a\right) ,$ $k_{y}=n_{y}\,\left( \pi /a\right) $
with $n_{x},\,n_{y}=0,\pm 1,\pm 2...,$ Eq. (\ref{qum}) can be rewritten as

\begin{equation}
\varepsilon \left( \frac{n_{x}}{n_{0}},\,\frac{n_{y}}{n_{0}}\right) =\frac{B%
}{2}\left[ 2-\cos \left( \pi \frac{n_{x}}{n_{0}}\right) -\cos \left( \pi 
\frac{n_{y}}{n_{0}}\right) \right] ,  \label{ram}
\end{equation}
where $n_{0}=a/\sigma .\;$To find the isoenergetic contours on the $%
n_{x},\,n_{y}$ plane, the initial values $n_{x}/n_{0}=\theta $ and $%
n_{y}/n_{0}=0$ are chosen corresponding to energy

\begin{equation}
\varepsilon \left( \theta \right) =\frac{B}{2}\left[ 1-\cos \left( \pi
\theta \right) \right] .  \label{rem}
\end{equation}%
For $\theta =1,$ we have $\varepsilon \left( 1\right) =B$ which is the
maximum value of kinetic energy on the isoenergetic contours. This means
that $B$ represents the actual band-width. Then, by keeping $\varepsilon
\left( n_{x}/n_{0},\,n_{y}/n_{0}\right) =\varepsilon \left( \theta \right) ,$
$n_{y}/n_{0}$ is evaluated as a function of $n_{x}/n_{0}$ for various values
of $\theta $ by means of a numerical procedure which also finds the area
enclosed in the contours. Owing to electron spin, twice this area represents
the number $N_{s}$ of states of energy less than $\varepsilon \left( \theta
\right) .$ For $\theta =1,$ the contour is a square of half-diagonal $n_{0}$
and area $2n_{0}^{2}$ (see Figure 1). Thus, for $\theta =1,$ $N_{s}$ assumes
its maximum value $N_{s\,0}=4n_{0}^{2}.$ On the other hand on the CuO$_{2}$
planes each mesh of area $\sigma ^{2}/2$ holds one electron, so that the
overall number of electrons is $2n_{0}^{2}.$ This means that the band is
half-filled. For $N_{s}/N_{s\,0}=0.5,$ we find $\theta =0.59$. It follows,
from Eq. (\ref{rem}), $\varepsilon _{F}=0.64B\ $or, by identifying the Fermi
energy with the chemical potential, $\mu =0.64B.$ Utilizing the previously
mentioned numerical data, the density of states per unit cell at the Fermi
level is found to be

\begin{equation}
\Omega _{F}=\frac{1}{N}\,\frac{dN_{s}}{d\varepsilon _{F}}=\frac{1}{N}\,\frac{
N_{s\,0}}{B},  \label{rim}
\end{equation}
which, taking into account that $2n_{0}^{2}=N,$ leads to the simple
relationship

\begin{equation}
\mu \Omega _{F}=1.28.  \label{rom}
\end{equation}%
It is to be pointed out that this result holds in general independently of
band-width $B$, that is, of the actual values of overlap and exchange
integrals.

\FRAME{ftbphFU}{4.427in}{3.5864in}{0pt}{\Qcb{Isoenergetic contours for
electrons running on CuO$_{2}$ planes. The square contour for $n_{x}/n_{0}=1$
corresponds to band-width energy $B$, the one for $n_{x}/n_{0}=0.59$ to the
Fermi energy $\protect\varepsilon _{F}$.}}{}{fig3x.png}{\special{language
"Scientific Word";type "GRAPHIC";maintain-aspect-ratio TRUE;display
"USEDEF";valid_file "F";width 4.427in;height 3.5864in;depth
0pt;original-width 11.4588in;original-height 9.2708in;cropleft "0";croptop
"1";cropright "1";cropbottom "0";filename 'Fig3X.png';file-properties
"XNPEU";}}

\bigskip \bigskip \bigskip

\bigskip

\bigskip

\bigskip {\Large Appendix C - Conservation of momentum in fermion systems}

Hamiltonian (\ref{bim}) conserves momenta of the interacting electrons. To
realize how this occurs, we recall that in degenerate fermion systems, by
letting\ $n_{r}$ be the number of particles on the r-th level and $\omega
_{r}$ the level degeneracy, we have, at zero temperature, $n_{r}=\omega _{r}$
below Fermi level and $n_{r}=0$ above it. Consequently, taking into account
that fluctuation of $n_{r}$ is

\begin{equation}
\delta \,n_{r}=\sqrt{n_{r}\left( 1-n_{r}/\omega _{r}\right) },  \label{xiz}
\end{equation}%
no fluctuation is allowed $\left[ 53\right] $.$\;$For the same reason, no
scattering is possible either (\footnote{%
) For $T>0$ K, states\ in the Fermi level region are only partially
occupied.\ Thus, in this region scattering processes can take place. But it
is wrong \ to ascribe superconductivity to these scatterings because \ when $%
\ T$ \ goes to zero scatterings are shut off while the\ superconductig
energy gap $\Delta _{0}$ attains its maximum amplitude. Fermi level appears
in Eq. (\ref{don}) since $\Delta $ is small with respect to chemical
potential $\mu $, as explained in Eq. (\ref{dem}).}).\ This is a direct
consequence of the Pauli principle. A clear evidence about this property of
degenerate fermion systems is provided by the nuclear matter.\ In fact,
nuclei can be described both by the liquid drop model, in which nucleons
strongly interact among themselves, and by the shell model, in which
nucleons behave as particles moving freely in a potential well.\ This
confirms that in degenerate fermion systems interacting particles maintain
unchanged their momenta.

We must therefore verify that Hamiltonian (\ref{bim}) really does not allow
for scattering processes. We have

\begin{equation}
\widehat{H}_{pair}=-W_{P}\sum_{k,\text{ }k^{\prime }}\widehat{\alpha }%
_{k^{\prime }a}^{+}\widehat{\alpha }_{k^{\prime }b}^{+}\widehat{\alpha }%
_{kb}^{{}}\widehat{\alpha }_{ka}^{{}}.  \label{xlz}
\end{equation}%
Terms with $k\neq k^{\prime }$ change pairs with momenta $k,a$ and $k,b$ in
pairs with momenta $k^{\prime },a$ and $k^{\prime },b$. But each term with $%
k=p$ and $k^{\prime }=q$ is associated with a term\ with $k=q$ and $%
k^{\prime }=p$. These terms represent opposite scattering processes.
Omitting\ factor $-W_{P}$, their contribution in $\widehat{H}_{pair}$ is

\begin{equation}
\widehat{S}_{pq}=\widehat{\alpha }_{qa}^{+}\widehat{\alpha }_{qb}^{+}%
\widehat{\alpha }_{pb}^{{}}\widehat{\alpha }_{pa}^{{}}+\widehat{\alpha }%
_{pa}^{+}\widehat{\alpha }_{pb}^{+}\widehat{\alpha }_{qb}^{{}}\widehat{%
\alpha }_{qa}^{{}}.  \label{xmz}
\end{equation}%
Taking into account the Jordan and Wigner anticommutation rules, operator $%
\widehat{S}_{pq}$\ can be rewritten in the ordered form

\begin{equation}
\widehat{S}_{pq}=\widehat{\alpha }_{qa}^{+}\widehat{\alpha }_{pa}^{{}}%
\widehat{\alpha }_{qb}^{+}\widehat{\alpha }_{pb}^{{}}+\widehat{\alpha }%
_{qa}^{{}}\widehat{\alpha }_{pa}^{+}\widehat{\alpha }_{qb}^{{}}\widehat{%
\alpha }_{pb}^{+}.  \label{xnz}
\end{equation}%
Operators $\widehat{\alpha }\ $and $\widehat{\alpha }^{+}$ and fermion
states $\left| 0\right\rangle \ $and $\left| 1\right\rangle $\ can be
represented by the matrices

\begin{equation}
\widehat{\alpha }=\left| 
\begin{array}{cc}
0 & 1 \\ 
0 & 0%
\end{array}%
\right| ,\quad \widehat{\alpha }^{+}=\left| 
\begin{array}{cc}
0 & 0 \\ 
1 & 0%
\end{array}%
\right| ,\quad \left| 0\right\rangle =\left| 
\begin{array}{c}
1 \\ 
0%
\end{array}%
\right| ,\quad \left| 1\right\rangle =\left| 
\begin{array}{c}
0 \\ 
1%
\end{array}%
\right| .  \label{xoz}
\end{equation}%
We have, indeed,

\begin{equation}
\widehat{\alpha }\left| 1\right\rangle =\left| 0\right\rangle ,\quad 
\widehat{\alpha }^{+}\left| 0\right\rangle =\left| 1\right\rangle ,\quad 
\widehat{\alpha }\left| 0\right\rangle =0,\quad \widehat{\alpha }^{+}\left|
1\right\rangle =0.  \label{xpz}
\end{equation}%
We consider, moreover, the hermitian operator

\begin{equation}
\widehat{\chi }=\widehat{\alpha }+\widehat{\alpha }^{+}=\left| 
\begin{array}{cc}
0 & 1 \\ 
1 & 0%
\end{array}%
\right|  \label{xqz}
\end{equation}%
which interchanges the fermion states, that is,

\begin{equation}
\widehat{\chi }\left\vert 0\right\rangle =\left\vert 1\right\rangle ,\quad 
\widehat{\chi }\left\vert 1\right\rangle =\left\vert 0\right\rangle .
\label{xrz}
\end{equation}%
In this way,\ operator $\widehat{S}_{pq}$ can be written as the sum of two 8$%
\times $8 matrices, each showing on the diagonal four 2$\times $2 matrices
for the $\widehat{\alpha }\ $and $\widehat{\alpha }^{+}$ operators
corresponding to $qa$, $pa$, $qb$, $pb$ states. So, by taking into account
Eq. (\ref{xqz}), we have

\begin{equation}
\widehat{S}_{pq}=\widehat{\chi }_{qa}\widehat{\chi }_{pa}\widehat{\chi }_{qb}%
\widehat{\chi }_{pb.}  \label{xsz}
\end{equation}%
By considering the four fermion state $\left| 1,1,1,1\right\rangle =\left|
1\right\rangle _{qa}\left| 1\right\rangle _{pa}\left| 1\right\rangle
_{qb}\left| 1\right\rangle _{pb},$ we obtain

\begin{equation}
\left\langle 1,1,1,1\right\vert \widehat{S}_{pq}\left\vert
1,1,1,1\right\rangle =\left\langle 1,1,1,1\right\vert \left.
0,0,0,0\right\rangle =0,  \label{xtz}
\end{equation}%
that is, the expectation value of $\widehat{S}_{pq}$ vanishes.\ The same
result is found\ in general for all fermion states. Consequently, no
scattering process is really accounted for in $\widehat{H}_{pair}$
Hamiltonian. Only non-scattering terms with $k=k^{\prime }$ give a
contribution. This means that electron momenta are conserved in Lewis
pairs.\ This argument applies unchanged to Cooper and Majorana Hamiltonians
in which scattering terms can be eliminated in the same way. The lack of
effective scattering processes entails that momentum of fermions is
conserved for all the pairing mechanisms considered in Table 4.\bigskip

\medskip

{\Large References }

$\left[ 1\right] $ G.P.\ Collins, Scientific American, August (2000) pag.
22; Scientific American, May (2004) pag. 15.

$\left[ 2\right] $ P\ Brovetto, V.\ Maxia and M.\ Salis, Nuovo Cimento D 19
(1997) 73.

$\left[ 3\right] $ P.\ Brovetto, V.\ Maxia and M.\ Salis, Eur. Phys. J. B 17
(2000) 85$.$

$\left[ 4\right] $ P.\ Brovetto, V.\ Maxia and M.\ Salis, Eur. Phys. J. B 21
(2001) 331.

$\left[ 5\right] $ P\ Brovetto, V.\ Maxia and M.\ Salis, J. Supercond. 14
(2001) 717.

$\left[ 6\right] $ J.F.\ Schooley, W.R. Hosler and L.\ Cohen Marvin, Phys.
Rev. Lett. 12 (1964) 474.

$\left[ 7\right] $ A.W.\ Sleight, J.L.\ Gillson and P.E.\ Bierstedt, Solid
State Comm. 17 (1975) 27.

$\left[ 8\right] $ R.J.\ Cava, B.\ Batlogg J.J.\ Kajewski, R.\ Favrow, L.W.\
Rupp, A.E.\ White, K.\ Short, W.F.\ Pecks and T.\ Kometan, Nature 334 (1988)
814.

$\left[ 9\right] \;$S.\ M.\ Kazakov, C.\ Chaillout, P.\ Bordet, J.J.\
Capponi, M.\ Nunez Regueiro, A.\ Rysak\S , J.L.\ Tholence, P.G.\ Radaelli,
S.N.\ Putilin and E.V.\ Antipov, Nature 390 (1997) 148.

$\left[ 10\right] $ J.G. Bednorz and K.A. Muller, Z.\ Phys.\ B 64 (1986) 189.

$\left[ 11\right] $\ M.K. Wu, J.R. Ashburn, C.J.\ Torng, P.H.\ Hor, R.L.\
Meng, L.\ Gao, Z.J. Huang, Y.Q.\ Wang, and C.W. Chu, Phys. Rev. Lett. 58
(1987) 908.

$\left[ 12\right] $\ C.C.\ Torardi, M.A.\ Subramanian, J.C.\ Calabrese, J.\
Gopalakrishnan, T.R.\ Askew, R.B.\ Flippen, K.J.\ Morrisey, U.\ Chowdhry and
A.W.\ Sleight, Science 240 (1988) 631.

$\left[ 13\right] $\ F.S. Galasso, Perovskites and High Tc Superconductors,
Gordon and Breach Science Publishers, New York, 1990.\ 

$\left[ 14\right] $ M.G.\ Smith, A.\ Manthiram, J.\ Zhou, J.B.\ Goodenough
and J.T.\ Markert, Nature 351 (1991) 549.

$\left[ 15\right] $ J.H.\ Sch\"{o}n, M. Dorget, F.C. Beuran, X.Z.\ Zu, E.
Arushanov, C.\ Deville Cavellin and M.\ Lagu\"{e}s, Nature 414 (2001) 434.

$\left[ 16\right] $ D.\ Jerome, in Studies of High Temperature
Superconductors, Vol. 6, 113 edited by A. Narkilar, Nova Science Publishers,
New York, 1990.

$\left[ 17\right] $ M.J.\ Rosseinsky, A.P.\ Ramirez, S.H.\ Glarum, D.W.\
Murphy, R.C.\ Haddon, A.F.\ Hebard, T.T.M.\ Palstra, A.R.\ Kortan, S.\ M.\
Zaurak, and A.W. Makhija, Phys. Rev. Lett.\ 66 (1991) 2830.

$\left[ 18\right] $ J.H.\ Sch\"{o}n, Ch. Kloc, B.\ Batlogg, Nature 408
(2002) 549.

$\left[ 19\right] $ G.N. Lewis, J. A. C. S. 38 (1916) 762.

$\left[ 20\right] $ W.\ Heitler and F.\ London, Z.\ Phys. 44 (1927) 445.

$\left[ 21\right] $ W.\ Heisemberg, Z.\ Phys. 49 (1928) 619.

$\left[ 22\right] $ E.\ Majorana, Z.\ Phys. 82, 137 (1933).

$\left[ 23\right] $ E$.$ Fermi, Nuclear Physics, The University of Chicago
Press, 1950, Ch.\ VI.

$\left[ 24\right] $ J.\ Bardeen, L.N.\ Cooper and J.R.\ Schrieffer, Phys.
Rev.\ 108 (1957) 1175.

$\left[ 25\right] $ N.N$.$ Bogolyubov, Z. Eksp. Teor. Fiz.\ 34 (1958) 58,
73; Nuovo Cimento 7 (1958) 794.

$\left[ 26\right] \;$A.\ Bohr, B.R.\ Mottelson and D. Pines, Phys. Rev. 110
(1958) 936.

$\left[ 27\right] \;$S.\ T.\ Belyaev, Mat.\ Fys.\ Medd.\ Dan.\ Vid. Selsk.
31 (1959) no. 11; Sov. Phys. JEPT\ 9 (1958) 23.

$\left[ 28\right] $ L.P.\ Gor'kov, Sov. Phys. JETP 7 (1958) 505.

$\left[ 29\right] $ A.\ I.\ Alekseev, Sov. Phys. Usp. 4 (1961) 1, 23.

$\left[ 30\right] $\ C.P.\ Poole, Jr., H. A. Farach and R.J.\ Creswick,
Superconductivity, Academic Press, New York, 1995.

$\left[ 31\right] $ L. Pauling and E.B.\ Wilson, Introduction to Quantum
Mechanics, McGraw-Hill, New York, 1935, \S\ 43.

$\left[ 32\right] $ L.\ D.\ Landau and E.\ M.\ Lifshitz, Statistical
Physics, Pergamon Press, Oxford, 1969, \S\ 80.\ 

$\left[ 33\right] $ G.\ Grosso, G.\ Pastori Parravicini, Solid State
Physics, Academic Press, New York, 2000.

$\left[ 34\right] $ J.P.\ Rice, J. Giapintzakis, D.M.\ Ginsberg and J.M.\
Mochel, Phys Rev. B 44 (1991) 10158.

$\left[ 35\right] $ K.A.\ Muller, Physica C 185-189 (1991) 3.

$\left[ 36\right] \;$I.K.$\;$Gopalakrishnan, in High Temperature
Superconductivity, edited by K.B.\ Garg and S.M.\ Bose, Narosa Publishing
House, New Delhi, 1998.

$\left[ 37\right] $ L.\ Soderholm, K.\ Zhany, D.G.\ Hinks, M.A.\ Beno, J.D.\
Jorgensen, C.U.\ Segre and Ivan K.\ Schuller, Nature 328 (1987) 604.

$\left[ 38\right] $\ H.A.\ Blackstead, J.D.\ Dow, Solid State Communications
115 (2000) 137.

$\left[ 39\right] $ D.P.\ Norton, D.H.\ Lowndes, B.C.\ Sales, J.D.\ Budai,
B.C.\ Chakoumakos and H.R.\ Kerchner, Phys. Rev. Lett.\ 66 (1991) 1537.

$\left[ 40\right] $ K.\ Takenaka, Y.\ Imanaka, K.\ Tamasaku, T.\ Ito and S.\
Uchida, Phys. Rev.\ B 46 (1992) 5833.

$\left[ 41\right] $ I.D.\ Brown, Acta Crystallogr. B 33 (1977) 1305.

$\left[ 42\right] $ I.D.\ Brown, Phys. Chem. Minerals 15 (1987) 30.

$\left[ 43\right] $ I.D.\ Brown and D.\ Altermat, Acta Crystallogr. B 41
(1985) 244.

$\left[ 44\right] $ R.J.\ Cava, A.W.\ Hewat, B.\ Batlogg, M.\ Marezio, K.M.\
Rabe, J.J.\ Kajewski, W.F.\ Pecks Jr and L.W. Rupp Jr, Physica C 165 (1990)
419.

$\left[ 45\right] $ J.\ Ruvald, Supercond. Sci. Technol. 9 (1996) 905.

$\left[ 46\right] $ C.P.\ Poole, Jr., T.\ Datta and H.A.\ Farach, J.\
Supercond.\ 2 (1989) 369.

$\left[ 47\right] $\ Mun-Seong Kim, Jae-Hyuk Choi and Sung-Ik Lee, Phys.
Rev. B 63 (2001) 324.

$\left[ 48\right] \;$G.C.\ Pimentel and R.D. Spratley, Chemical bonding
clarified through quantum mechanics, Holden Day, San Francisco, 1970, Ch. 7.

$\left[ 49\right] $ Z. Te\v{s}anovi\'{c}, Phys. Rev. B 36 (1987) 2364.

$\left[ 50\right] \;\ $T.M.\ Mishonov, A.A.\ Donkov, R.K.\ Koleva and E.S.\
Penev, Bulgarian J.\ Phys. 24 (1997) 114.

$\left[ 51\right] $ \ T.M.\ Mishonov, J.O.\ Indekeu and E.S.\ Penev, Int.\
J.\ Mod. Phys.\ B 16 (2002) 4577.

$\left[ 52\right] $\ J.L.\ Tholence, B. Souletie, O.\ Laborde, J.J.\
Capponi, C. Chaillout and M.\ Marezio, Phys. Lett. A (1994) 184.

$\left[ 53\right] $ R.H. Fowler, Statistical Mechanics, Cambridge University
Press, 1955, \S\ 20.21.

\bigskip

\end{document}